\input harvmac
\input epsf.tex
\input labeldefs.tmp
\input mssymb.tex

\writedefs

\overfullrule=0mm

\newcount\figno
\figno=0
\def\fig#1#2#3{
\par\begingroup\parindent=0pt\leftskip=1cm\rightskip=1cm\parindent=0pt
\baselineskip=11pt
\global\advance\figno by 1
\midinsert
\epsfxsize=#3
\centerline{\epsfbox{#2}}
\vskip 12pt
{\ninepoint%
{\bf Fig.~\the\figno:} {#1}}\par
\endinsert\endgroup\par
}
\def\figlabel#1{\xdef#1{\the\figno}%
\writedef{#1\leftbracket \the\figno}%
}
\def\encadremath#1{\vbox{\hrule\hbox{\vrule\kern8pt\vbox{\kern8pt
\hbox{$\displaystyle #1$}\kern8pt}
\kern8pt\vrule}\hrule}}
\def\omit#1{}

\def\conf{configuration}

\def\nind{\par\noindent}

\def\pre#1{ ({\tt
#1})}

\def\IR{\relax{\rm I\kern-.18em R}}
\font\cmss=cmss10 \font\cmsss=cmss10 at 7pt

\font\cmss=cmss10 \font\cmsss=cmss10 at 7pt
\def\IZ{\relax\ifmmode\mathchoice
{\hbox{\cmss Z\kern-.4em Z}}{\hbox{\cmss Z\kern-.4em Z}}
{\lower.9pt\hbox{\cmsss Z\kern-.4em Z}}
{\lower1.2pt\hbox{\cmsss Z\kern-.4em Z}}\else{\cmss Z\kern-.4em Z}\fi}
\def\IN{\relax{\rm I\kern-.18em N}}
\def\IZ{\Bbb{Z}}\def\IN{\Bbb{N}}
\def\Kr{Krattenthaler}
%
\lref\BIBLE{D. Bressoud, {\sl Proofs and confirmations. The story of the alternating
sign matrix conjecture}, Cambridge University Press (1999).}
\lref\RS{A.V. Razumov and Yu.G. Stroganov, 
{\sl Combinatorial nature
of ground state vector of O(1) loop model},
{\it Theor. Math. Phys.} 
{\bf 138} (2004) 333-337; {\it Teor. Mat. Fiz.} 138 (2004) 395-400 \pre{math.CO/0104216}.}
\lref\RSb{P. A. Pearce, V. Rittenberg and J. de Gier, 
{\sl Critical Q=1 Potts Model and Temperley-Lieb Stochastic Processes}
\pre{cond-mat/0108051}
\semi 
A.V. Razumov and Yu.G. Stroganov,
{\sl O(1) loop model with different boundary conditions 
and symmetry classes of alternating-sign matrices}
\pre{cond-mat/0108103}.}

\lref\MNosc{S. Mitra and B. Nienhuis, {\sl 
Osculating random walks on cylinders}, in
{\it Discrete random walks}, 
DRW'03, C. Banderier and
C. Krattenthaler edrs, Discrete Mathematics and Computer Science
Proceedings AC (2003) 259-264\pre{math-ph/0312036}.} 
\lref\MNdGB{S. Mitra, B. Nienhuis, J. de Gier and M.T. Batchelor,
{\sl Exact expressions for correlations in the ground state 
of the dense $O(1)$ loop model}\pre{cond-math/0401245}.}
\lref\BdGN{M.T. Batchelor, J. de Gier and B. Nienhuis,
{\sl The quantum symmetric XXZ chain at $\Delta=-1/2$, alternating sign matrices and 
plane partitions},
{\it J. Phys.} A {\bf 34} (2001) L265-L270
\pre{cond-mat/0101385}.}
\lref\dG{J.~de~Gier, {\sl Loops, matchings and alternating-sign matrices}
\pre{math.CO/0211285}.}
\lref\Wie{B. Wieland, {\it  A large dihedral symmetry of the set of
alternating-sign matrices}, 
Electron. J. Combin. {\bf 7} (2000) R37\pre{math.CO/0006234}.}
\lref\LGV{B. Lindstr\"om, {\it On the vector representations of
induced matroids}, Bull. London Math. Soc. {\bf 5} (1973)
85-90\semi
I. M. Gessel and X. Viennot, {\it Binomial determinants, paths and
hook formulae}, Adv. Math. { \bf 58} (1985) 300-321. }
\lref\DFZJZ{P.~Di~Francesco, P.~Zinn-Justin and J.-B.~Zuber,
{\sl A Bijection between classes of Fully Packed Loops and Plane Partitions},
{\it Electron. J. Combi.} to appear \pre{math.CO/0311220}.}
\lref\twostep{J. Grassberger, A. King and P. Tirao, {\sl On the homology of
free 2-step nilpotent Lie algebras}, {\it J. Algebra} {\bf  254}, 213-225 (2002).}
{
\catcode`\~=12%
\lref\Sloane{On-Line Encyclopedia of Integer Sequences,\hfill\break
{\tt http://www.research.att.com/$\sim$njas/sequences/Seis.html}%
.}
}
\lref\Krtr{
 M. Ciucu, C.\Kr, T. Eisenk\"olbl and D. Zare, 
{\sl Enumeration of lozenge tilings of hexagons with a central triangular hole}, 
{\it J. Combin. Theory Ser.} A 95 (2001), 251-334
\pre{math.CO/9910053}\semi
C. \Kr, {\sl Descending plane partitions and rhombus tilings of a hexagon with triangular hole}
\pre{math.CO/0310188}.}
\lref\Krdet{C. \Kr,  {\sl Advanced determinant calculus}, 
{\it S\'eminaire Lotharingien Combin.} 42 (``The Andrews Festschrift") (1999), Article B42q
\pre{math.CO/9902004}.}
\lref\Krcorn{M. Ciucu and C. \Kr, {\sl Enumeration of lozenge tilings of hexagons with cut-off corners}, 
{\it J. Combin. Theory} Ser. A 100 (2002), 201-231 \pre{math.CO/0104058}.}
%
%
\lref\DFZ{P.~Di~Francesco and J.-B.~Zuber, {\sl On FPL 
configurations with four sets of nested arches}, 
JSTAT (2004) P06005 \pre{cond-mat/0403268}.}
\lref\AvM{M.~Adler and P.~van~Moerbeke, {\sl Virasoro action on Schur function expansions, skew Young tableaux and random walks}
\pre{math.PR/0309202}.}
\lref\Kratt{F. Caselli and C.~\Kr, {\sl Proof of two conjectures of
Zuber on fully packed loop configurations}, {J. Combin. Theory
Ser.} {\bf A 108} (2004), 123-146, \pre{math.CO/0312217}.  }
%
%
%
\Title{\vbox{\hbox{SPhT-T04/120}\hbox{LPTHE-04-23}}}
{\vbox{
\centerline{Determinant Formulae for some Tiling Problems}
\medskip
\centerline{and Application to Fully Packed Loops}
}}
\bigskip
\centerline{P.~Di~Francesco \footnote{${}^\#$}{
Service de Physique Th\'eorique de Saclay,
CEA/DSM/SPhT, URA 2306 du CNRS,
C.E.A.-Saclay, F-91191 Gif sur Yvette Cedex, France},
 P. Zinn-Justin \footnote{${}^\star$}
{LIFR--MIIP, Independent University, 
119002, Bolshoy Vlasyevskiy Pereulok 11, Moscow, Russia   and
Laboratoire de Physique Th\'eorique et Mod\`eles Statistiques, UMR 8626 du CNRS,
Universit\'e Paris-Sud, B\^atiment 100,  F-91405 Orsay Cedex, France
}
and J.-B. Zuber ${}^\#$\footnote{${}^\flat$}{
LPTHE, Tour 24, Universit\'e Paris 6, 75231 Paris Cedex 05},}
\medskip


\bigskip\bigskip\bigskip
\noindent
We present a number of determinant formulae for the number of tilings
of various  domains in relation with Alternating Sign Matrix
and Fully Packed Loop enumeration. 

\bigskip\bigskip\bigskip

\font\eightrm=cmr8
\centerline{\eightrm AMS Subject Classification (2000): Primary 05A19; Secondary 52C20, 82B20}

\Date{09/2004}

%
%

\newsec{Introduction}
Determinants appear naturally in physics when one studies systems of free fermions: their wave
functions (Slater wave function), as well as their grand canonical partition function,
can be expressed as determinants. These statements have exact discrete counterparts.
In particular, discrete dynamics can be formulated in terms of transfer
matrices (instead of a Hamiltonian): these are familiar objects of statistical mechanics,
which from a combinatorial point of view simply count the number of ways to go from a given
initial configuration to a given final configuration. Here we are more specifically
interested in models on two-dimensional lattices, in which the role of free fermions
is played by non-intersecting paths; the analogous determinantal expressions 
can then be derived from
the so-called Lindstr\"om--Gessel--Viennot (LGV) formula~\LGV. 
In this paper we intend to show how these determinantal techniques 
can be applied to some combinatorial problems, which all amount to
the enumeration of certain types of rhombus tilings.


The methods we present here are fairly general 
and make use of various
standard combinatorial objects. Young diagrams appear 
as a way to encode locations of paths crossing
a given straight line;
 furthermore, the determinants involved often turn out to be Schur functions
$s_Y(x)$ where, beside the Young diagram $Y$, 
appears the set of parameters $x=\{x_1,x_2,\ldots\}$ which
encodes the dynamics (here we are mostly concerned with simple counting, in which case $x_i=1$).
Also note that in the context of free fermions, Schur functions are natural building blocks for tau-functions
of classical integrable hierarchies, see for example \AvM\ 
for such a connection and further relations to matrix 
integrals.

The paper is organized as follows. In section 2, we introduce the
basic methods and tools which are required for our computations -- the LGV formula, and
the definition of the elementary transfer matrices from which all can
be  built -- and revisit as an example the MacMahon formula.
In section 3, we provide some further examples
of enumeration of rhombus tilings of various domains. 
In section 4, we show how similar 
considerations also enable us to 
count fully packed loop (FPL) configurations with given connectivities
and various symmetries. We conclude in section 5 with some open
issues, including
a discussion of the asymptotic enumeration of FPL configurations.

\newsec{Basic ingredients: corner transfer matrices and the LGV formula}
\subsec{The LGV formula}
Here and in the following we consider lattice paths whose oriented steps
connect nearest neighboring vertices of the two-dimensional integer
lattice $\IZ^2$, and with only two directions allowed say along the vectors
$(-1,0)$ and $(0,1)$. 
The LGV formula allows to express the number ${\cal P}(\{S\},\{E\})$ of configurations of
$n$ non-intersecting lattice paths with, say, starting points $S_1,S_2,\ldots,S_n$
and endpoints $E_1,E_2,\ldots,E_n$
in $\IZ^2$, as the determinant of the matrix whose element $(i,j)$
is the number ${\cal P}(S_i,E_j)$ of lattice paths from $S_i$ to $E_j$, namely
\eqn\lgvformula{ {\cal P}(\{S\},\{E\})=\det\big({\cal P}(S_i,E_j)
\big)_{1\leq i,j\leq N}\ . }
For reasons that will become obvious later on, the matrix
with entries ${\cal P}(S_i,E_j)$ will be called {\it  transfer
matrix}\/
for the corresponding path counting problem.

\subsec{The matrices W and T}
\fig{The transfer matrix $T$ is expressed as a product $W W^t$.}{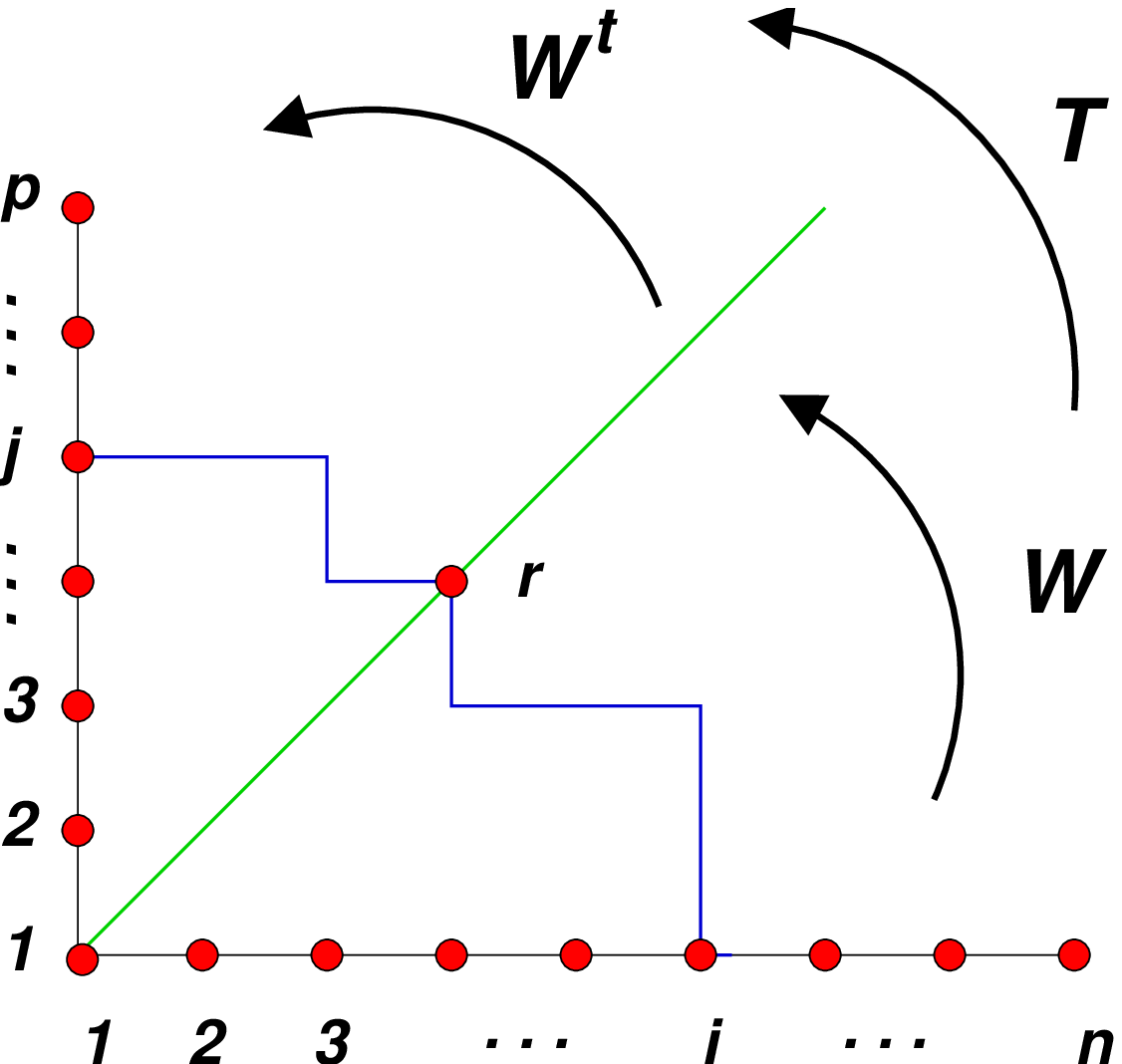}{6.cm}
\figlabel\tmatrix
In this note we will mainly be dealing with lattice paths with starting and endpoints
occupying consecutive positions along two lines that intersect. The most basic situations 
are depicted in Fig.~\tmatrix. The most fundamental situation is that of (ordered)
starting points $S_a=(i_a,1)$, $a=1,2,\ldots,N$, and endpoints $E_a=(r_a,r_a)$, $a=1,2,\ldots,N$, namely
with paths across a corner of angle $45^{\circ}$. 
The corresponding ``corner'' transfer matrix
$W$ has the entries 
\eqn\entryW{ W_{i,r}\equiv {\cal P}\big((i,1),(r,r)\big)={i-1\choose r-1} }
expressing that $r-1$ vertical steps must be chosen among a total of $i-1$.
The action of $W$ is represented in the lower corner of Fig.~\tmatrix.
The LGV formula allows us to write
\eqn\appliW{ {\cal P}(\{(i_a,1)\}_{a=1,\ldots,N},\{(r_b,r_b)\}_{b=1,\ldots,N})=\det\left( W_{i_a,r_b}
\right)_{1\leq a,b\leq N}\ .}
Upon reflecting the picture and exchanging the roles of starting and endpoints, we easily
find that $W^t$ is the corner transfer matrix for the situation with starting
points $(r_a,r_a)$ and endpoints $(1,i_b)$. We deduce that the corner transfer matrix
for the situation $S_a=(i_a,1)$ and $E_b=(1,j_b)$ is nothing 
else than 
$T=W W^t$. We easily find
that
\eqn\Telem{ T_{i,j}\equiv {\cal P}\big((i,1),(1,j)\big)=\sum_{r\geq 1}
W_{i,r}W_{j,r}={i+j-2\choose i-1} }
which eventually expresses that $i-1$ horizontal steps must be taken among a total of $i+j-2$.
Similarly, the LGV formula gives
\eqn\appliT{ {\cal P}(\{(i_a,1)\}_{a=1,\ldots,N},\{(1,j_b)\}_{b=1,\ldots,N})=\det\left( T_{i_a,j_b}
\right)_{1\leq a,b\leq N}\ .}

Note that in both \appliW\ and \appliT\ the result is simply the minor determinant corresponding
to a specific choice of $N$ rows and columns of the matrices $W$ or $T$, taken of 
sufficiently large size.

\subsec{Rhombus tiling of a hexagon: the MacMahon formula revisited}
As a first application of the above, let us evaluate the total number of rhombus tilings
of a hexagon of size $a\times b\times c$, using the three elementary rhombi 
of size $1\times 1$.

\fig{A sample rhombus tiling of a hexagon of size $a\times b\times c$ (left). We have
indicated in red the De Bruijn lines following the two types of rhombi
shown in the oval. We have deformed them so as to obtain a set of $a$
non-intersecting lattice paths (right).}{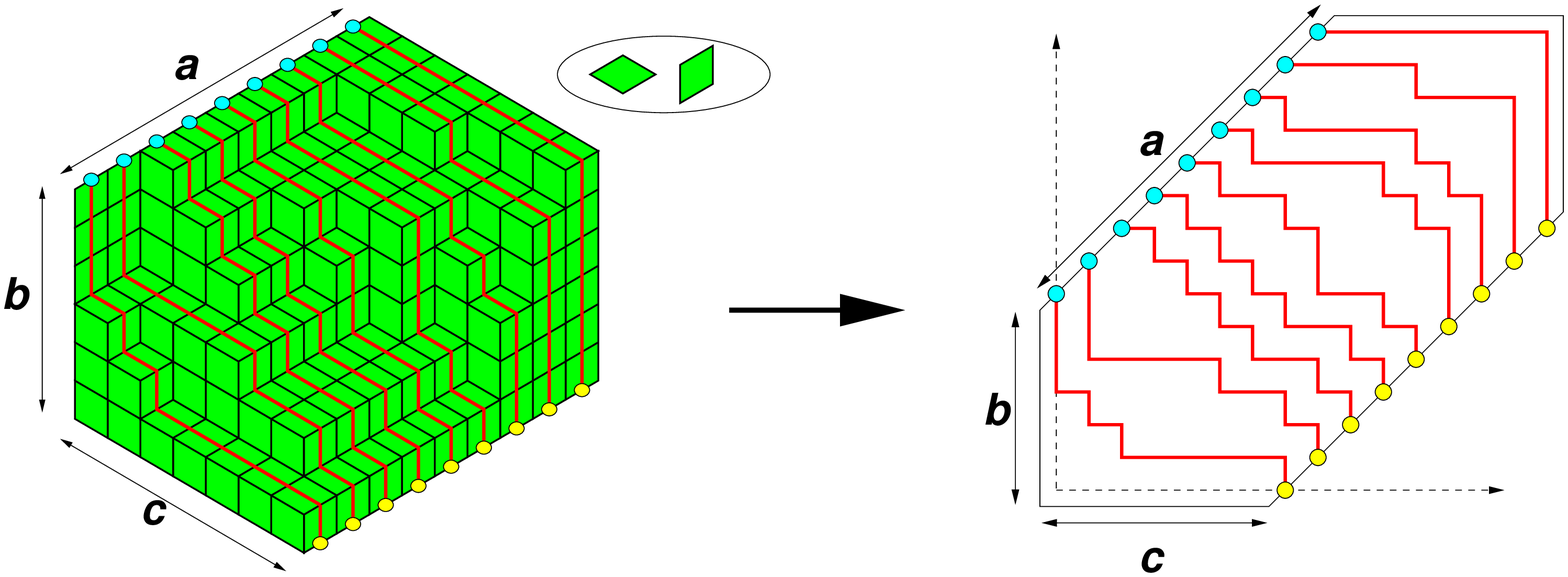}{15.cm}
\figlabel\macma

The standard approach consists of introducing
so-called De Bruijn lines that follow chains of consecutive rhombi of two of the three types.
In Fig.~\macma, we have represented the $a$ De Bruijn lines connecting consecutive
points along the two sides of length $a$ of the hexagon. 
Upon slightly deforming the rhombi, as indicated in Fig.~\macma,
these lines form a set of $a$ non-intersecting lattice paths, with starting points
$S_i=(c+i-1,i)$, $i=1,2,\ldots,a$ and endpoints $E_j=(j,b+j-1)$, $j=1,2,\ldots,a$.
Moreover, rhombus tilings of the hexagon are in bijection with such non-intersecting lattice
path configurations. The total number of such configurations reads, according to the LGV
formula:
\eqn\mac{ N(a,b,c)=\det\left( H_{b,c}(a)_{i,j}\right)_{1\leq i,j \leq a} }
where we have introduced the ``parallel'' transfer matrix 
$H_{b,c}$ of size $a\times a$, with entries
\eqn\entryH{H_{b,c}(a)_{i,j}={b+c\choose b+j-i}\ .}
Note that Eq.~\mac\ expresses $N(a,b,c)$ as the Schur function associated to the $b\times a$ rectangular 
Young diagram and to parameters $x_1=\cdots=x_{b+c}=1$ (i.e.\ dimension as a $GL(b+c)$ representation). 
Similar occurrences will be discussed in more detail below.

By line and column manipulations, this determinant may be explicitly computed
to yield the celebrated MacMahon formula
\eqn\macmahon{ N(a,b,c)=\prod_{i=1}^a\prod_{j=1}^b \prod_{k=1}^c
{i+j+k-1\over i+j+k-2}\ . }

\fig{Using $T$ to enumerate rhombus tilings of a hexagon with a given
number of winding De Bruijn loops. The first picture shows a sample rhombus tiling, and its
decomposition into three parallelograms. In each parallelogram, the De Bruijn
lines follow sequences of the two rhombi displayed in the ovals. In the second picture, 
we only retain the De Bruijn lines: within each parallelogram,
these are clearly generated by the corner transfer
matrix $T$ with the appropriate shape.}{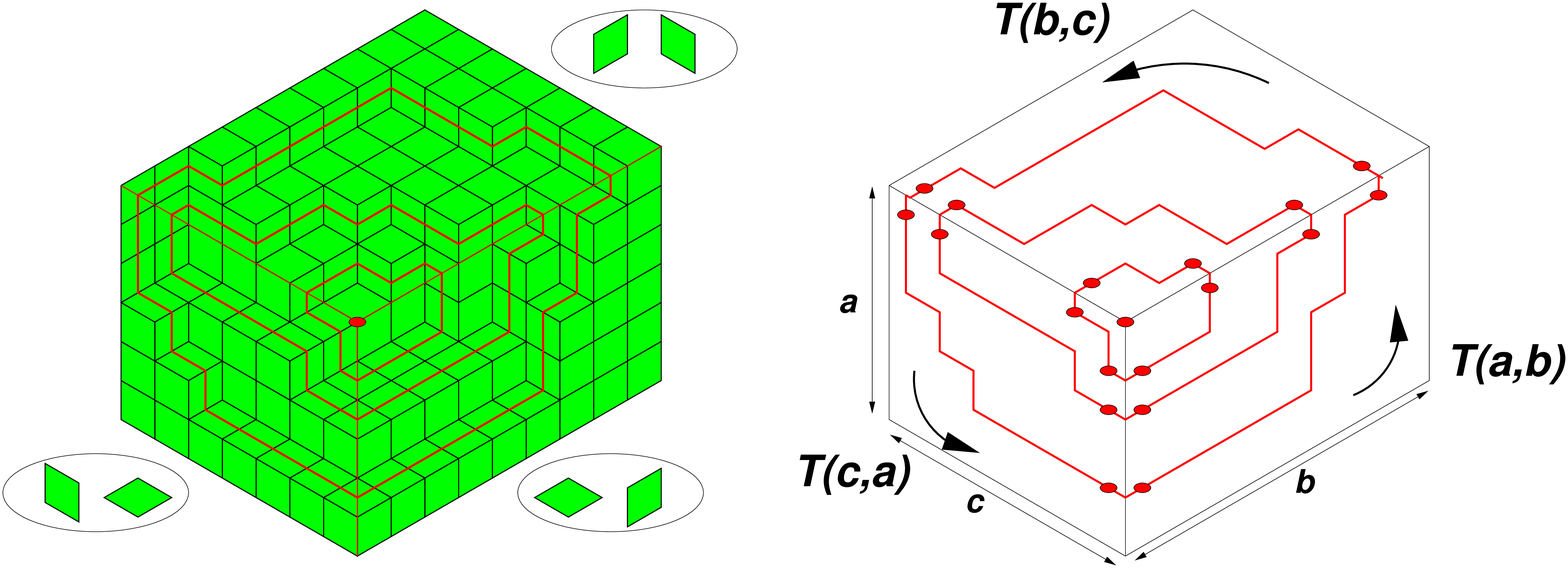}{15cm}
\figlabel\hexa

We may alternatively decompose the hexagon into three parallelograms (there are exactly two ways of 
doing this, pick one), meeting at a ``central point" (see Fig.~\hexa). We now follow particular
De Bruijn lines joining the inner edges of these parallellograms {\it without ever exiting}\/ the hexagon.
These are sequences
of two of the three tiles, the pairs differing in the three
parallellograms. These lines actually form loops that wind around the central point.
Note finally that the problem of counting configurations of lines with fixed ends within 
each of the three parallelograms uses a corner transfer matrix $T$ of appropriate size.
In Fig.~\hexa, we have denoted by $T(x,y)$ the rectangular matrix with entries \Telem, for  
$i=1,2,\ldots,x$ and $j=1,2,\ldots,y$.
It is now easy to write the number $N_d(a,b,c)$ of configurations of exactly $d$ non-intersecting
loops winding around the central point:
\eqn\dnumber{\eqalign{ N_d(a,b,c)&= 
\sum_{\matrix{1\leq & a_1<\cdots<a_d & \leq a\cr
1\leq & b_1<\cdots<b_d & \leq b \cr
1\leq & c_1<\cdots<c_d & \leq c}  }
\det_{1\leq r,s\leq d}\left( T_{a_r,b_j}T_{b_j,c_k}T_{c_k,a_s}\right)\cr
&=\sum_{1\leq a_1<\cdots<a_d\leq a} \det_{1\leq r,s\leq d}\left( 
\big(T(a,b)T(b,c)T(c,a)\big)_{a_r,a_s}\right)\cr
&=\det_{a\times a}( I +\mu\, T(a,b)T(b,c)T(c,a))\vert_{\mu^d}\cr}}
where the last line uses the decomposition formula for the determinant of the sum of two
matrices in terms of their minors, and we have to extract the $\mu^d$ coefficient to 
get the sum over all diagonal minors of size $d\times d$. This gives a refinement 
of the MacMahon formula.
We recover the total number of tilings of the hexagon by summing over $d$,
or equivalently by taking $\mu=1$ in the last determinant, namely
\eqn\total{ N(a,b,c)=\sum_{d=0}^{{\rm min}(a,b,c)}
N_d(a,b,c)=\det(I+T(a,b)T(b,c)T(c,a))\ .}
Note that the symmetry under the permutations
of $a,b,c$ is manifest here, as opposed to
eq. \mac. In the case $a=b=c$, denoting $T(a)\equiv T(a,a)$, we have
\eqn\ototal{ N(a,a,a)=\det(I+T(a)^3)=\det(I+T(a))\det(I+\omega T(a))\det(I+\omega^2 T(a))}
where $\omega=e^{2i\pi/3}$. Nice formulas \refs{\Krtr,\Krdet} 
happen to be known for the characteristic
polynomial of $T(a)$ precisely at sixth roots of unity, and allow for yet another
expression for the MacMahon formula in a cube, via \ototal.

\fig{Decomposition of the transfer matrix $H_{b,c}(a)$ for De Bruijn lines of a 
$a\times b\times c$ hexagon which
allows to keep track of paths passing to the left or right of the ``center"
(red dot). We give an explicit representation of the matrices $W$ and $Z_b$.}{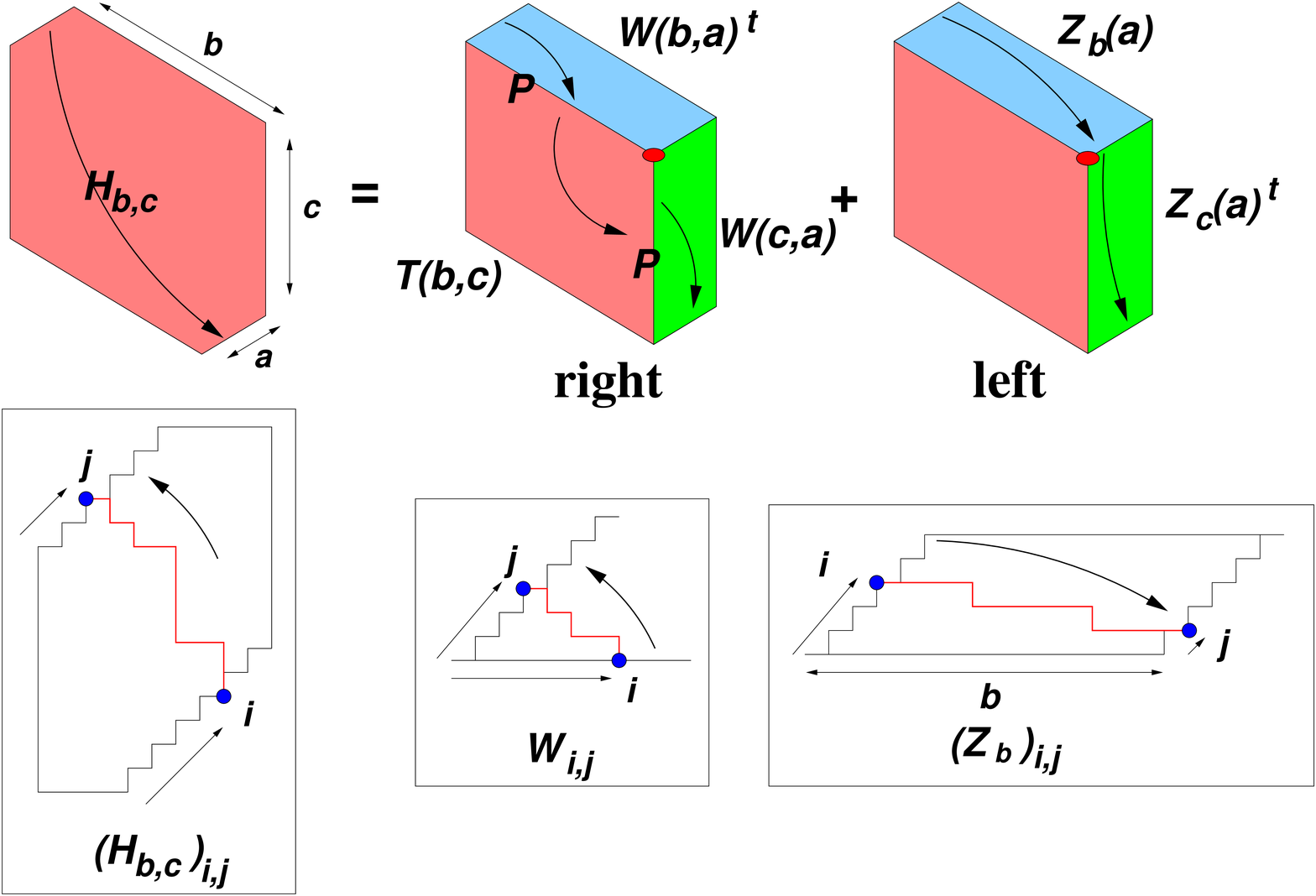}{13.cm}
\figlabel\leftright

It easy
to relate this picture to the former, with a set of $a$ standard
De Bruijn lines connecting the opposite sides of length $a$ in the hexagon.
Indeed consider the intersection of the standard De Bruijn lines, as well of winding lines,
with the line
parallel to the $a$ sides that goes through the center: the location of the lines
is precisely the same to the right of the center, whereas it is complementary to the left.
Therefore $N_d(a,b,c)$ also counts the tiling configurations 
{\it with exactly 
$d$ of these lines passing to the right of the ``center" of the hexagon}.
This is also a consequence of the following relation between transfer 
matrices, pictorially proved  in Fig.~\leftright:
\eqn\relatmat{ H_{b,c}(a)= W(b,a)^t P_b T(b,c) P_c W(c,a) + Z_b(a)Z_c(a)^t }
in which the $(ij)$ entry of the first term counts the number of paths
{\it passing to the right of the center}\/ connecting the i-th entry
point to the j-th exit one, while the second term yields the contribution of
paths passing to the {\it left}\/ of the center.
In \relatmat, as usual 
the argument corresponds to the size of the matrix ($(a)$ for a square $a\times a$ matrix,
$(a,b)$ for a rectangular $a\times b$ matrix), and the matrices have
the following  entries, see Fig. \leftright:
\eqn\entrHPTZ{ (H_{b,c})_{ij}={b+c\choose c+i-j}, \qquad W_{ij}={i-1\choose j-1},
\qquad (P_b)_{ij}=\delta_{j,b+1-i},\qquad (Z_b)_{ij}={b\choose i-j}\ . }
Note that the matrix $P$ induces a reflection, needed whenever passing from a corner
to another 
one pointing in the opposite direction.
We finally identify
\eqn\tmhex{\det_{a\times a}( I +\mu \,T(a,b)T(b,c)T(c,a))=\det_{a\times a}(Z_b(a)Z_c(a)^t+\mu 
V(b,a)^t P_b T(b,c) P_c V(c,a)) }
by noticing that, after factoring out $Z Z^t$ whose determinant is 1, 
the two corresponding matrices  are conjugate of one-another. 

\newsec{Rhombus tilings of various domains}
\subsec{Lozenge with glued sides}
\fig{Computing the numbers of tilings of a lozenge of side a
glued along two of its consecutive edges, and with exactly d loops
winding around its conic singularity.}{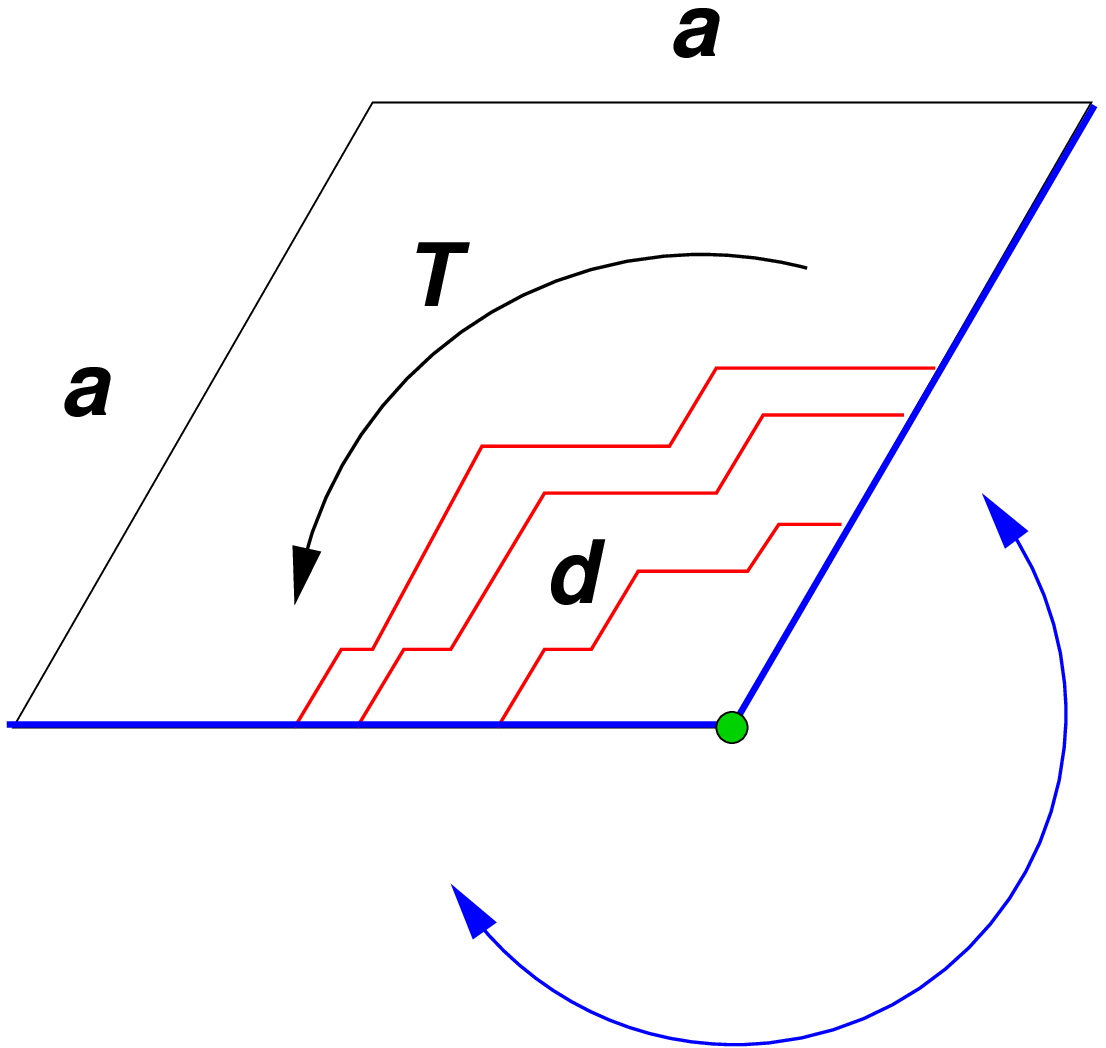}{5.cm}
\figlabel\twosideglu

$T$ may be used to compute the numbers $N_d^{L}(a)$ of tilings of a lozenge of side $a$
glued along two of its consecutive edges (see Fig.~\twosideglu), and
on which exactly $d$ De Bruijn loops wind
around the tip of the cone. We simply have
\eqn\simply{ N_d^L(a)=\det_{a\times a} (I+\mu \,T(a))\vert_{\mu^d}\ . }

Disregarding the $d$ dependence, we find the numbers
\eqn\exone{ N^L(a)=\det_{a\times a}\left( I+T(a)\right)= \prod_{n=0}^{a-1} {(3n+2)(3n)!n!\over
(2n)!(2n+1)!}\ . }
These read for $a=1,2,3,\ldots$
\eqn\tilglu{ 2,5,20,132,1452,26741,826540,42939620, \ldots}
This is by construction the number of cyclically symmetric plane partitions in a cube 
$a\times a\times a$,
obtained by decomposing a regular hexagon of size $a\times a\times a$ into three identically
tiled parallelograms, with suitable identifications of boundaries
(see section 4.3 for another interpretation of these numbers).

\subsec{Half-hexagon with glued sides}
\fig{How to use the corner transfer matrices $W$ and $W^t$ to generate the 
rhombus tilings of a half-hexagon with glued sides. The gluing is indicated by the arrows.
The half-hexagon is decomposed into three triangles, within which the De Bruijn lines
are generated by $W$ or $W^t$.}{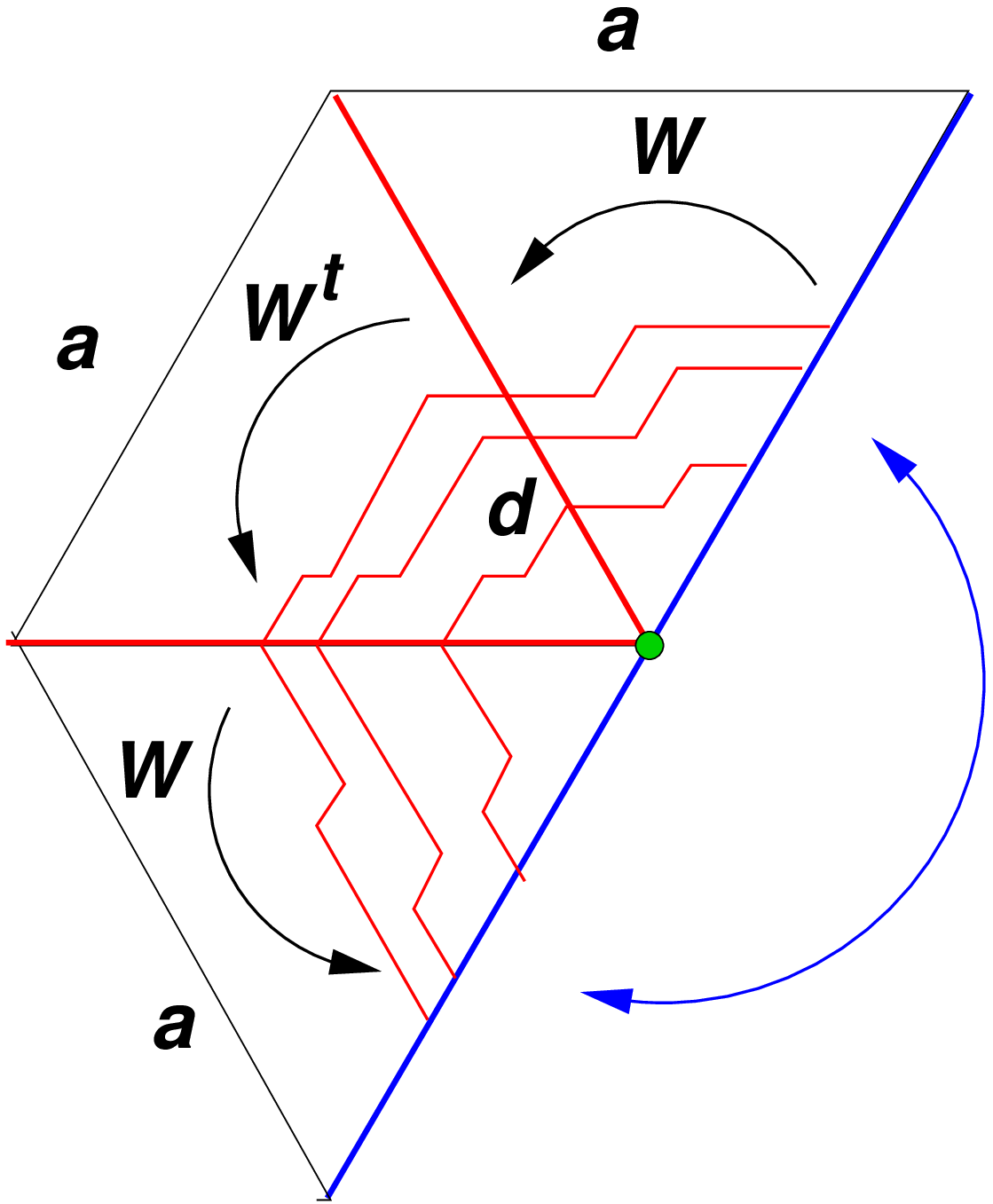}{5.cm}
\figlabel\hexaglu

The corner transfer matrix $W$ may be used to compute the numbers $N_d^{HH}(a)$ 
of tilings of a half-hexagon of side $a$
glued along two of its cut edges (see Fig.~\hexaglu), and on which 
exactly $d$ De Bruijn loops wind
around the tip of the cone. We have
\eqn\simplyhex{N_d^{HH}(a)=\det_{a\times a} \big(I+\mu\, W(a) W(a)^t
W(a)\big)\vert_{\mu^d}\ . }

Disregarding the $d$ dependence, we find the numbers
\eqn\extwo{ N^{HH}(a)=\det_{a\times a}\left( I+WW^tW\right)= \left\{\matrix{
2^n \left(\prod_{j=0}^{n-2} {(4j+3)!(j+1)!j!\over (2j+2)!((2j+1)!)^2}\right)^2
{(4n-1)!n!(n-1)!\over (2n)!((2n-1)!)^2 } & {\rm if} \ a=2n \cr
2^{n+1} \left(\prod_{j=0}^{n-1} {(4j+3)!(j+1)!j!\over (2j+2)!((2j+1)!)^2}\right)^2
& {\rm if} \ a=2n+1 \cr} \right. }
These numbers read for $a=1,2,3,\ldots$
\eqn\tilhalf{ 2,6,36,420,9800,4527650,41835024,7691667984\ldots}
Remarkably,
these turn out to match exactly {\it the total dimension of the homology of  free 2-step nilpotent
Lie algebras of rank $a$} \twostep\ 
(cf entry A078973  of the on-line encyclopedia of integer sequences \Sloane).

Let us give an 
explanation for this apparent coincidence, and present a
determinant formula for the Poincar\'e polynomial of 2-step nilpotent
Lie algebras of rank $a$. As in the case of the full hexagon, one can 
introduce the
standard De Bruijn lines; here there are $a$ lines which enter from the middle side of length $a$ and come out
from the side of length $2a$. Let us denote by $2a\ge n_1>\cdots>n_a>0$ the locations of their endpoints.
Then the LGV formula tells us that the number of such De Bruijn lines at fixed endpoints $n_i$ is:
\eqn\hhnumb{
N^{HH}_{(n)}(a)=\det {a \choose n_i-(a+1-j)}_{1\le i,j\le a}\ .
}

Now the irreducible representations of $GL(a)$ are indexed by
Young diagrams $Y=\{\lambda_1,\lambda_2,\ldots,\lambda_a\}$ with 
$\lambda_i$ boxes in the i-th row, and $\lambda_i\geq \lambda_{i+1}\geq 0$
for $i=1,2,\ldots,a-1$. The corresponding character, evaluated on
the class of matrices with eigenvalues $x=\{x_1,x_2,\ldots,x_a\}$ is
the Schur function, expressed via the Jacobi--Trudi formula as
$s_Y(x)=\det\big(h_{\lambda_i+j-i}(x)\big)_{1\leq i,j\leq a}=\det\big(h'_{\lambda'_i+j-i}(x)\big)_{1\le i,j\le a}$
where $h_m(x)$ are the complete symmetric functions, generated by
$\sum_{m\geq 0}h_m(x)t^m=\prod_{i=1}^a 1/(1-tx_i)$, similarly
$\sum_{m\geq 0}h'_m(x)t^m=\prod_{i=1}^a (1+t x_i)$, and
the $\lambda'_i$ denote the $\lambda$'s of the transposed Young diagram
$Y^T$ (reflected w.r.t.\ the first diagonal). The dimension
$\dim_Y$ of the representation $Y$ correponds to the character of the
identity class, with $x_1=x_2=\cdots=1$ (also denoted by $x=1$). Explicitly,
\eqn\gamb{
\dim_Y=s_Y(1)=\det {a+\lambda_i+j-i-1\choose \lambda_i+j-i}_{1\leq i,j\leq a}
=\det {a\choose \lambda_i'+j-i}_{1\leq i,j\leq a}\ .}
We recognize in Eq.~\hhnumb\ this dimension, 
provided we identify $n_i=\lambda'_i+a+1-i$.
Finally, the gluing of the two half-sides means
in the summation over the endpoints $n_i$, we must only include those such that the corresponding diagram
$Y$ is invariant by transposition. We thus find:
\eqn\simplyhexb{N^{HH}(a)=\sum_{Y=Y^T} \dim_Y}
which is precisely the quantity calculated in \twostep.

\fig{(a) Half-hexagon tilings with $d$ De Bruijn lines having fixed endpoints, with positions
$m_1,m_2,\ldots,m_d$ and $p_1,p_2,\ldots,p_d$
counted from the center. (b) The corresponding Young diagrams, with characteristics
$(m;p)$ in Frobenius notation. Boxes are counted from and include the 
diagonal, from left to right ($m$'s) and top to bottom ($p$'s).}{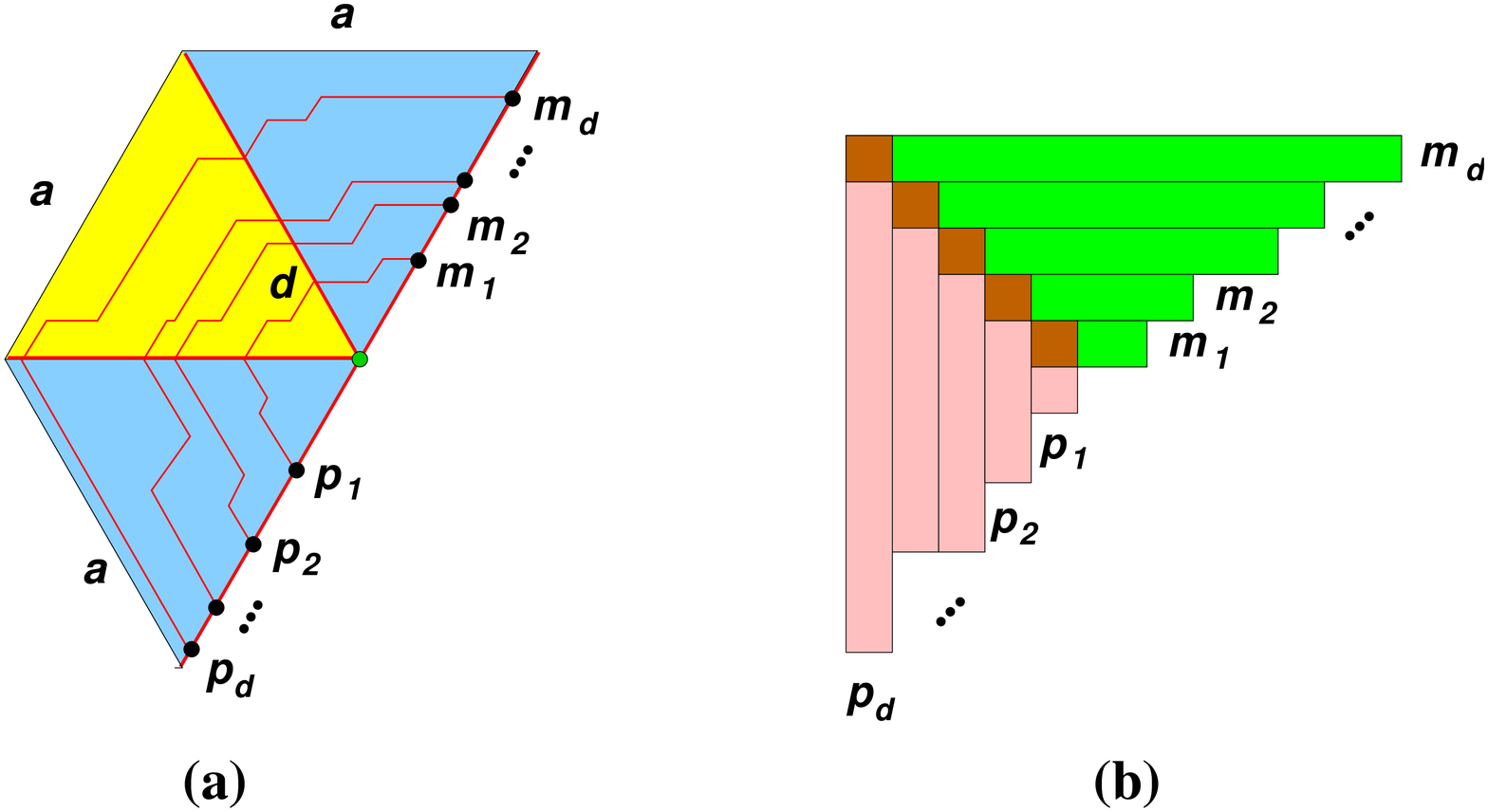}{10.cm}
\figlabel\hexatoschur

Note that the correspondence between the two approaches can be made for each term of the sum:
indeed the endpoints of the standard lines and of the winding lines are clearly in one-to-one
correspondence. If we record the positions
of endpoints of the winding lines on the (cut) boundary, counted from the center,
namely $(m;p)=\{ m_1,\ldots ,m_d; p_1,\ldots ,p_d\}$,
$1\leq m_1<m_2<\cdots<m_d\leq a$ from center to top and $1\leq p_1<p_2<\cdots<p_d\leq a$
from center to bottom in the case of $d$ winding lines, then the corresponding Young diagram $Y$
is given by Fig.~\hexatoschur.
Therefore, the total number
of tiling configurations $N^{HH}_{(m;p)}(a)$ of the half-hexagon with fixed 
endpoints $(m;p)$ of De Bruijn lines is also the dimension of $Y$:
\eqn\hhnum{ N^{HH}_{(m;p)}(a)=\det\big( (WW^tW)_{m_i,p_j}\big)_{1\leq
i,j\leq d}=\dim_Y \ .}

The Poincar\'e polynomial $P_a(u)$ generating the dimensions of the homology spaces
of a 2-step nilpotent Lie algebra of rank $a$ has been shown to read \twostep\
\eqn\poinca{ P_a(u)=\sum_{Y=Y^T } \dim_Y u^{|Y|} }
where the sum extends over the self-transposed Young diagrams,
 %
namely those for which the two characteristics $(m;p)$
are equal, and where $|Y|$ denotes the total number of boxes in $Y$. Note
that $|Y(m;m)|=\sum_{1\leq i\leq d} (2m_i-1)$. Introducing the
diagonal matrix $\theta$ with entries 
\eqn\thetentries{\theta_{ij}=\delta_{ij} u^{2i-1}, \qquad i,j=1,2,\ldots,a}
and using the relation \hhnum, we get
\eqn\homodim{ P_a(u)=\det_{a\times a}(I+\theta WW^tW)\ . }
The first few such polynomials read
\eqn\firfptwostep{ \eqalign{
P_1(u)&=1+u\cr
P_2(u)&=1 + 2 u + 2 u^3 + u^4\cr
P_3(u)&=1 + 3 u + 8 u^3 + 6 u^4 + 6 u^5 + 8u^6 + 3 u^8 + u^9\cr
P_4(u)&=1 + 4 u + 20 u^3 + 20 u^4 + 36 u^5 + 64 u^6 + 20 u^7 + 
90 u^8 + 20 u^9 + 64 u^{10} + 36 u^{11} \cr 
&+ 20 u^{12} + 20 u^{13} + 4 u^{15} + u^{16}\cr
P_5(u)&=1 + 5 u + 40 u^3 + 50 u^4 + 126 u^5 + 280 u^6 + 160 u^7 + 
765 u^8 + 245 u^9 + 1248 u^{10} + 720 u^{11} \cr
&+ 1260 u^{12} + 
1260 u^{13} + 720 u^{14} + 1248 u^{15} + 245 u^{16} + 765 u^{17} + 
160 u^{18} + 280 u^{19} + 126 u^{20} \cr
&+ 50 u^{21} + 40 u^{22} + 5 u^{24} + u^{25}\ .\cr}}
At $u=1$, we recover the formula \extwo, which indeed counts the total dimension
of the homology of a 2-step nilpotent Lie algebra of rank $a$.

\subsec{Hexagon with a triangular central hole}
\fig{The transfer matrix $T_m(a,b)$ is used to generate tilings
of a hexagon with a central triangular hole.}{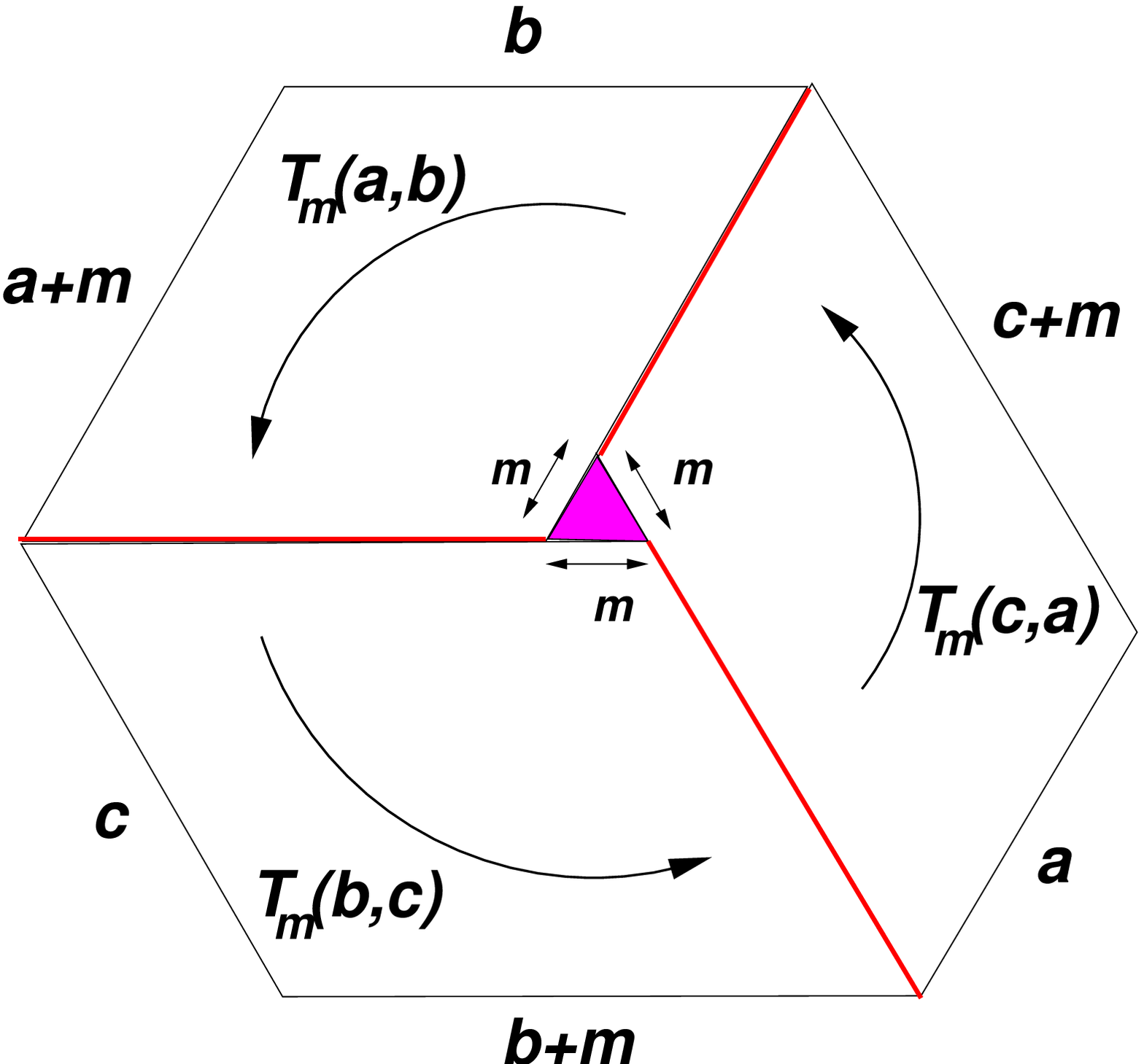}{7cm}
\figlabel\tmathole

A variant of $T$ may be used to generate the numbers $N_{d}(a,c+m,b,a+m,c,b+m)$ of
tilings of a hexagon with an equilateral central triangle of side $m$ removed
(see Fig.~\tmathole), and with $d$ loops winding around the hole. 
More precisely, let $T_m(a,b)$ denote the transfer matrix
of rectangular size $a\times b$ with entries
\eqn\newmat{ (T_m)_{ij}={m+i+j-2\choose j-1}}
with $i=1,2,\ldots,a$, $j=1,2,\ldots,b$.
The configurations are then generated by
\eqn\genew{ \det_{a\times a}\left(I+\mu \,T_m(a,b)T_m(b,c)T_m(c,a)\right)=
\sum_{d=0}^{{\rm min}(a,b,c)} \mu^d N_{d}(a,c+m,b,a+m,c,b+m)\ . }
See \Krtr\ for explicit expressions of some of  these determinants. 

\subsec{Hexagon with chopped-off corners}
\fig{A hexagon of shape $a\times b\times c$ with three chopped off corners
at distances $m,p,q$ from the origin (central point). We have represented
the relevant corner transfer matrix.}{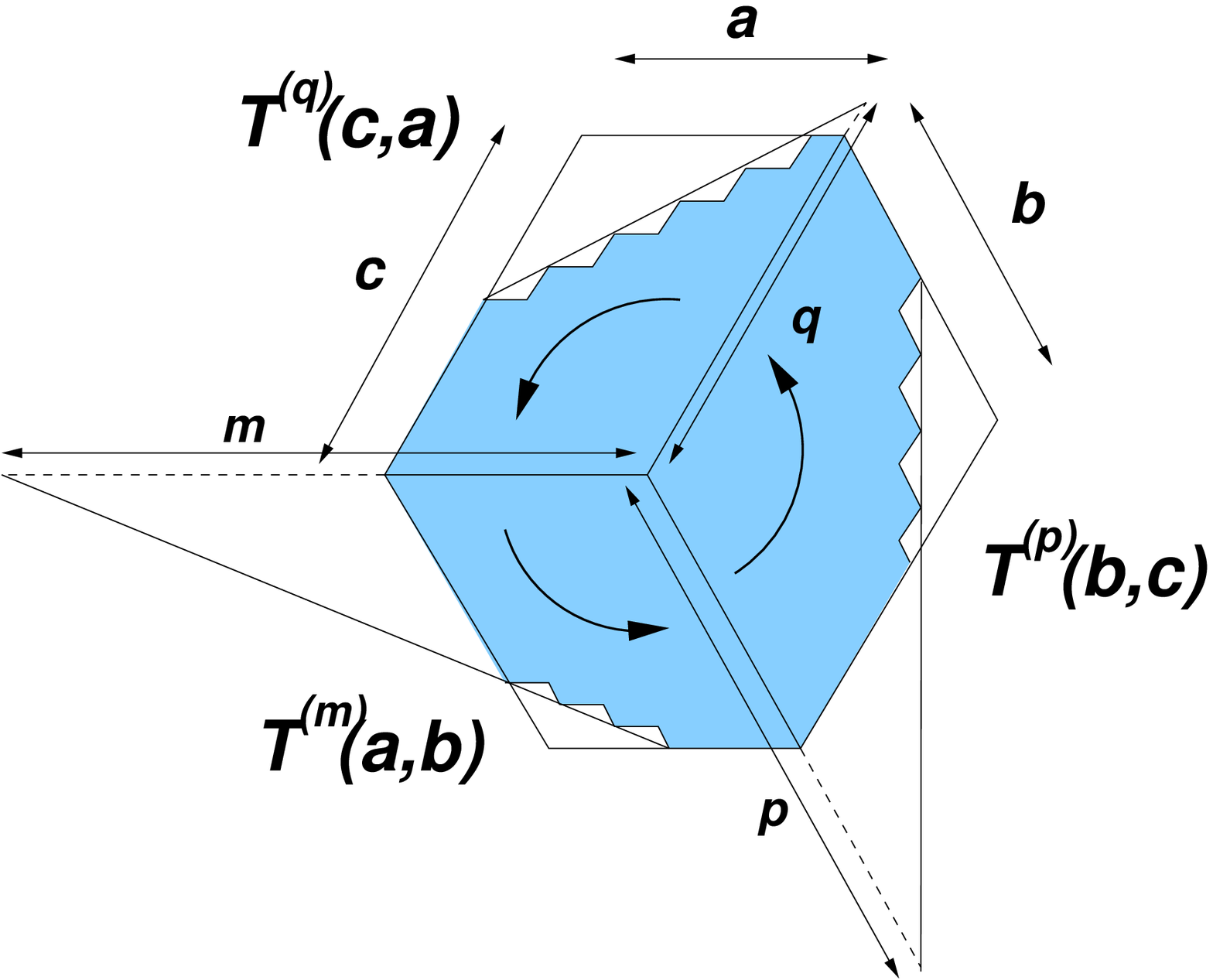}{8.cm}
\figlabel\chopped

We consider the tiling of a hexagon $a\times b\times c$ with three corners chopped
off as indicated in Fig.~\chopped, along line cuts at distances $m,p,q$
from the central point, with $m\geq {\rm max}(a,b)$,
$p\geq {\rm max}(b,c)$, and $q\geq {\rm max}(a,c)$. 
Let $N_d^{(m,p,q)}(a,b,c)$ denote the number of tilings of this domain with $d$ 
winding lines.  These are enumerated  using a restricted version of $T$
that incorporates a ``ceiling" at a given height not to be crossed. 
By the reflection principle,  the relevant truncation reads
\eqn\trucT{ T^{(m)}_{ij}={i+j-2\choose i-1}-{i+j-2\choose m}\ . }
We denote by $T^{(m)}(a,b)$ the corresponding rectangular matrix $a\times b$.
With this definition, we have
\eqn\gentrunc{ \det_{a\times a} \left( I +\mu\, T^{(m)}(a,b)T^{(p)}(b,c)T^{(q)}(c,a)
\right)\vert_{\mu^d}=N_d^{(m,p,q)}(a,b,c)\ . }

Note that a half-corner transfer matrix can still be introduced, namely a matrix 
$W^{(m)}$ with entries
\eqn\vmentr{ W^{(m)}_{ij}= {i-1\choose j-1}-{i-1\choose m+1-j} }
and such that $T^{(m)}(a,b)=W^{(m)}(a,c) W^{(m)}(b,c)^t$ for any $c\geq m/2$.

A particular case of \gentrunc\ corresponds to tilings of a
notched equilateral triangle, with $a=b=c$ and $m=p=q=a$.

\fig{Tilings of a hexagon with removed parallellograms. The relevant transfer
matrix $T$ has a block decomposition according to the arrows.}{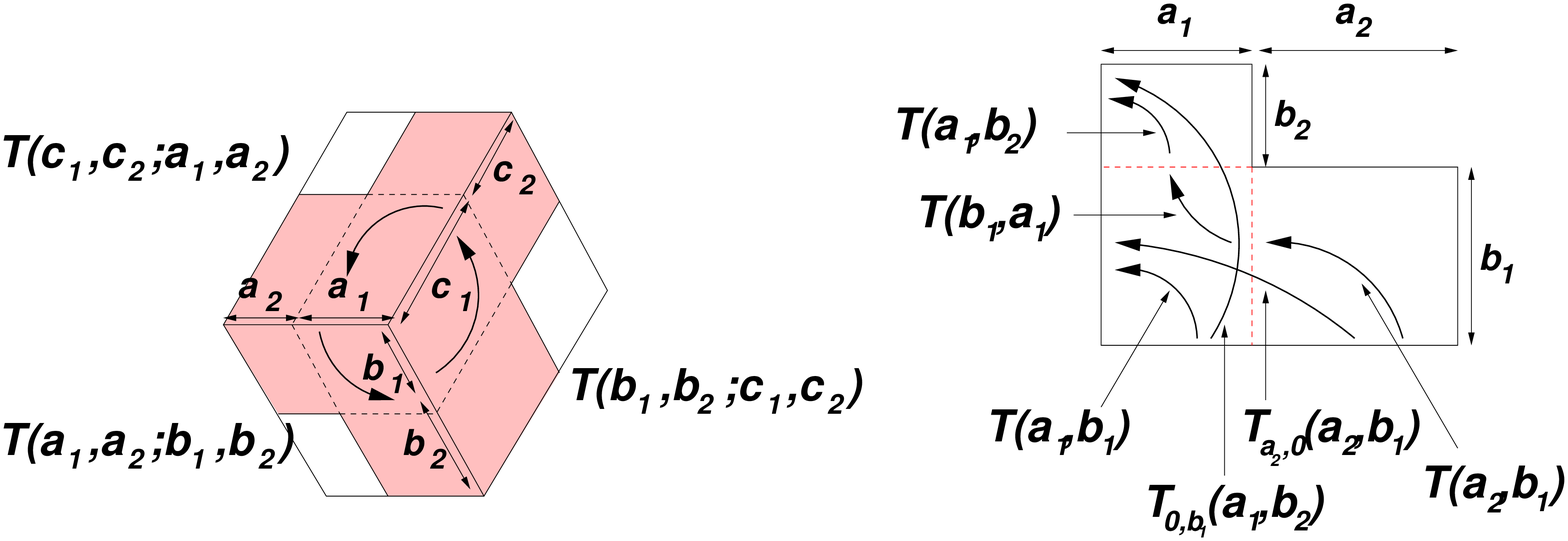}{13.cm}
\figlabel\broken

We may easily treat the case of rectangular
chopping as well, by suitably modifying $T$ to incorporate the corresponding
``broken ceiling" restriction indicated in Fig.~\broken. Let $N_d(a_1,a_2;b_1,b_2;c_1,c_2)$
denote the corresponding number of tilings with $d$ winding loops.
We have the transfer matrix
\eqn\transmatTbrok{ T(a_1,a_2;b_1,b_2)=\pmatrix{ T(a_1,b_1) & T_{0,b_1}(a_1,b_2)\cr
T_{a_1,0}(a_2,b_1) & T(a_2,b_1)T(b_1,a_1)T(a_1,b_2) } }
where $T_{m,p}(a,b)_{ij}={m+i+p+j-2\choose m+i-1}$, $1\leq i\leq a$, $1\leq j\leq b$,
and finally
\eqn\finbrok{\eqalign{ &N_d(a_1,a_2;b_1,b_2;c_1,c_2)=\cr
&\det_{(a_1+a_2)\times (a_1+a_2)}\left( I +\mu \,T(a_1,a_2;b_1,b_2)
T(b_1,b_2;c_1,c_2)T(c_1,c_2;a_1,a_2)\right)\vert_{\mu^d}\ . \cr}}


\newsec{Fully-Packed Loops and more tiling problems}
\subsec{Fully-Packed Loops, Alternating Sign Matrices and Plane Partitions}
The fully-packed loop (FPL) model plays a central role in the Razumov--Stroganov conjecture,
which attracted a lot of attention recently. In short, the FPL model is a statistical model
on a square grid of size $a\times a$ of the square lattice, in which edges may be occupied or
not, and with the constraint that exactly two incident edges are occupied at each vertex.
Moreover, one imposes alternating boundary conditions along the border of the grid, that 
every second edge
at the exterior of and perpendicular to the boundary is occupied. These occupied
external edges are labeled $1,2,\ldots,2a$. The FPL configurations are in bijection with
alternating sign matrices (ASM) of size $a\times a$, namely matrices with entries $\pm 1,0$
only, such that $1$ and $-1$ alternate and sum up to $1$ along each row and column.

The Razumov--Stroganov conjecture involves refined FPL numbers, according to the connectivity
of external edges. Indeed, from the definition of the model, the occupied external edges
are connected by pairs via chains of consecutive occupied edges, forming lines
(rather than loops) on the square grid, while closed loops may occupy some of the inner edges
of the grid. To summarize these connectivities, one generally uses the language of link patterns,
namely {\it planar}\/ permutations $\pi\in S_{2a}$ with only 2-cycles indicating the connected
edges. The Razumov--Stroganov
conjecture relates the numbers of FPL with fixed connectivities to the groundstate vector
of the O(1) loop model \refs{\RS,\BdGN}.

Rhombus tiling configurations of a hexagon are also called {\it plane partitions} (PP),
as they may be interpreted as the view in perspective from the $(1,1,1)$ direction
of a piling-up of unit cubes in the positive octant of the 3D lattice $\IZ^3$, with 
the constraint that gravity has the direction $-(1,1,1)$ and that only stable configurations
are retained. The total number of FPL or ASM of size $a\times a$ matches that of
so-called totally symmetric self-complementary plane partitions (TSSCPP)
of size $2a\times 2a\times 2a$, namely plane partitions
being maximally symmetric, {\it i.e.}\ under rotations of $2\pi/3$ and reflections w.r.t.\ medians,
as well as identical to their complement.
This result is one of the keystones of modern combinatorics and is beautifully 
described in the book \BIBLE. From this we simply retain that there should exist a natural 
bijection between the two sets of objects, still to be found to this day.

\subsec{FPL with fixed sets of nested lines}
In parallel to the FPL--ASM--TSSCPP relation, there exist bijections between special FPL configurations
and rhombus tilings of special domains. The simplest of these concerns the FPL of size $n\times n$
with three sets of nested lines, in numbers say $a,b,c$ with $a+b+c=n$, namely with a link pattern 
that connects $2a$ external edges say $1,2,\ldots,2a$ by symmetric pairs $(1,2a)(2,2a-1),\ldots,(a,a+1)$,
then the $2b$ next analogously and the $2c$ remaining analogously, thus forming three ``bundles"
of respectively $a,b,c$ connecting lines. These were shown to be in bijection with
the rhombus tilings of a hexagon of size $a\times b\times c$ \DFZJZ, and henceforth are enumerated
by the MacMahon formula \mac. 

\fig{The domain whose rhombus tilings match the $N(a,b,c,d)$ FPL with four sets of nested lines $a,b,c,d$.
It is a heptagonal domain with two sides glued, and with the indicated size. Moreover,
it must have exactly $d$ De Bruijn loops (thick lines) crossing the
gluing line.
The transfer matrices
are obtained by decomposing the domain into three parallelograms as indicated.}{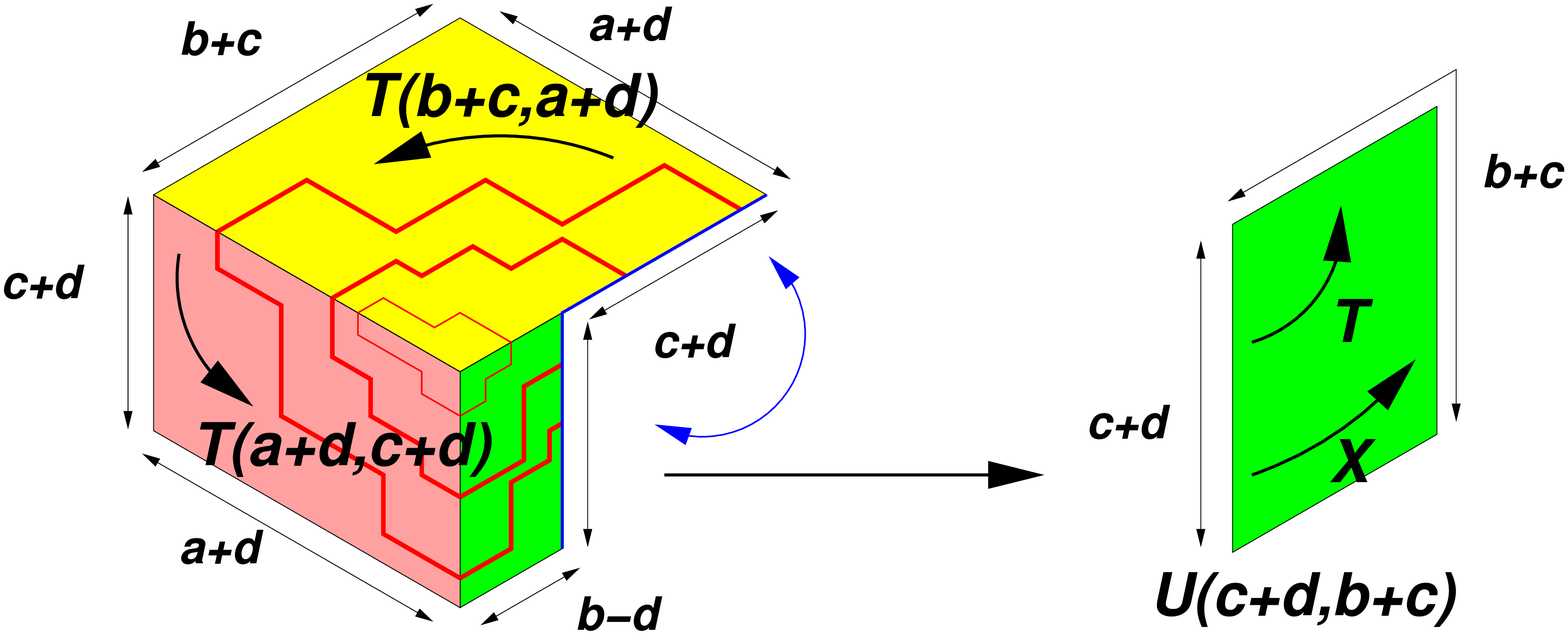}{13.cm}
\figlabel\fourlines

This result was improved in \DFZ\ so as to include the case of four sets of
nested lines, in numbers say $a,b,c,d$. The total number $N(a,b,c,d)$ of such FPL 
matches that of rhombus tilings
of the  domain of Fig.~\fourlines, which have exactly $d$  De Bruijn
loops crossing the gluing line of length $c+d$. 
This decomposition allows one to write the number as
\eqn\numfour{ N(a,b,c,d)=\det_{(b+c)\times (b+c)}
\big(I+T(b+c,a+d)T(a+d,c+d+e)U(c+d,b+c)\big)\vert_{\mu^d}}
where the transfer matrix $U(p,q)$ has the block form (see the right sketch of Fig.~\fourlines)
\eqn\blockU{ U(p,q)=\pmatrix{\vrule height 10pt depth3pt width0pt T(p,q-p) & \mu\,X_{q-p}(p)} }
where $(X_m)_{i,j}={m-1+i-j\choose i-j}$.

\fig{FPL configurations with four sets of nested arches separated
into two subsets by a fifth one.}{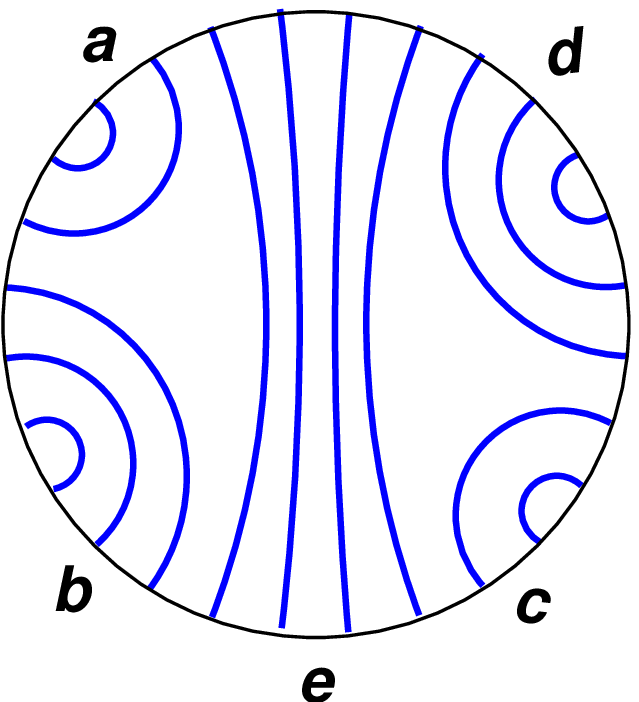}{3cm}\figlabel\cinqpaq

Finally, it is possible to extend this bijection to FPL with link patterns
with five sets of  nested lines as depicted on fig. \cinqpaq.
These FPL are easily
related to the rhombus tilings of the domain displayed in Fig.~\fivelines, which have
exactly $d$ De Bruijn loops going across the gluing line. 
The only difference with the case of four nested sets
of lines is that we have introduced a branch cut of length $e$ in the glued domain.
The numbers $N(a,b|e|c,d)$ of desired FPL reads
\eqn\fplnabcde{N(a,b|e|c,d)= \det_{(b+c)\times (b+c)}
\big(I+T(b+c,a+d)T(a+d,c+d+e)U^{(e)}(c+d+e,b+c)\big)\vert_{\mu^d}}
where the transfer matrix $U^{(e)}$ has the block form
\eqn\Ublock{ U^{(e)}(p,q)=\pmatrix{ T(p,q-p+e) & \matrix{ 0 \cr \mu\,X_{q-p+2e}(p-e) \cr}\cr}}
with a zero block of size $e\times (p-e)$. Note that all of the currently known
numbers of FPL configurations with prescribed connectivities (for which a proof exists), 
as in \refs{\DFZJZ,\DFZ,\Kratt}, are special cases of our pattern
$(a,b|e|c,d)$. 

\fig{The domain whose rhombus tilings match the $N(a,b|e|c,d)$
FPL with $a,b$ and $c,d$ nested lines separated by $e$. The nonagon
is glued along its edges of length $c+d$ as indicated by blue arrows. The red
segments of length $e$ are forbidden, and form a branch cut in the glued
domain. After decomposing the domain into three sub-domains,
we have indicated the transfer matrices needed for the enumeration. As before, 
we must have a total
of $d$ De Bruijn loops going through the glued cut of length
$c+d$.}{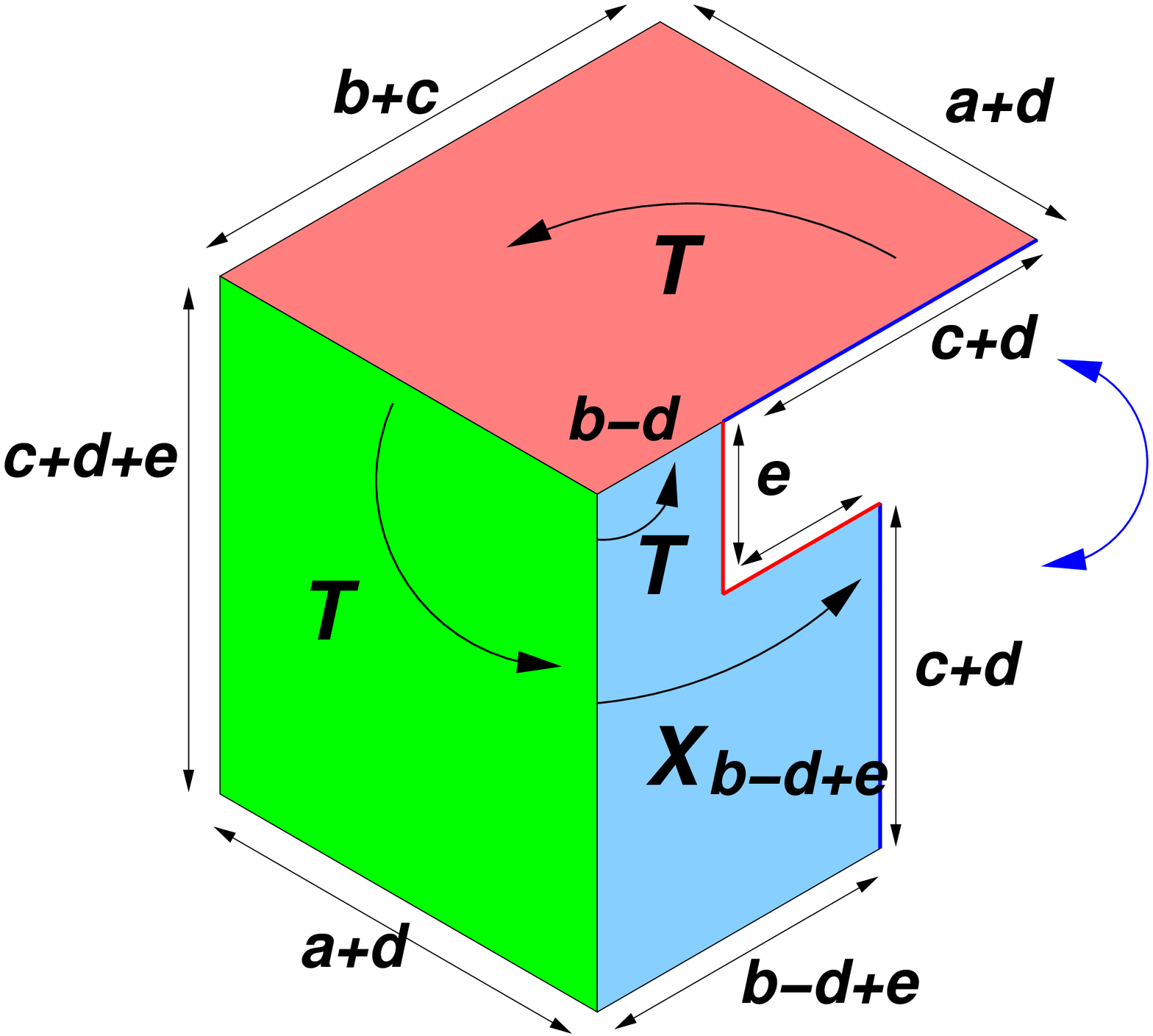}{7cm}
\figlabel\fivelines

\subsec{Other symmetry classes of FPL/ASM}
The Razumov--Stroganov has been generalized to symmetry classes of FPL:
the numbers of FPL with certain symmetries and
with given connectivity patterns turn out to be related to the ground state
vectors of the same $O(1)$ loop model, but with different boundary conditions \refs{\RSb,\dG}.
The counting of FPL with connectivity $(a,b|e|c,d)$ described above can be naturally
restricted to symmetry classes by imposing the corresponding symmetry on the
rhombus tiling, or equivalently by dividing out the domain by the symmetry.
Here we consider only the symmetry classes for which a Razumov--Stroganov type conjecture
is known.

\fig{Domain of rhombus tilings for $(a,b|e|a,b)$ HTSFPL.}{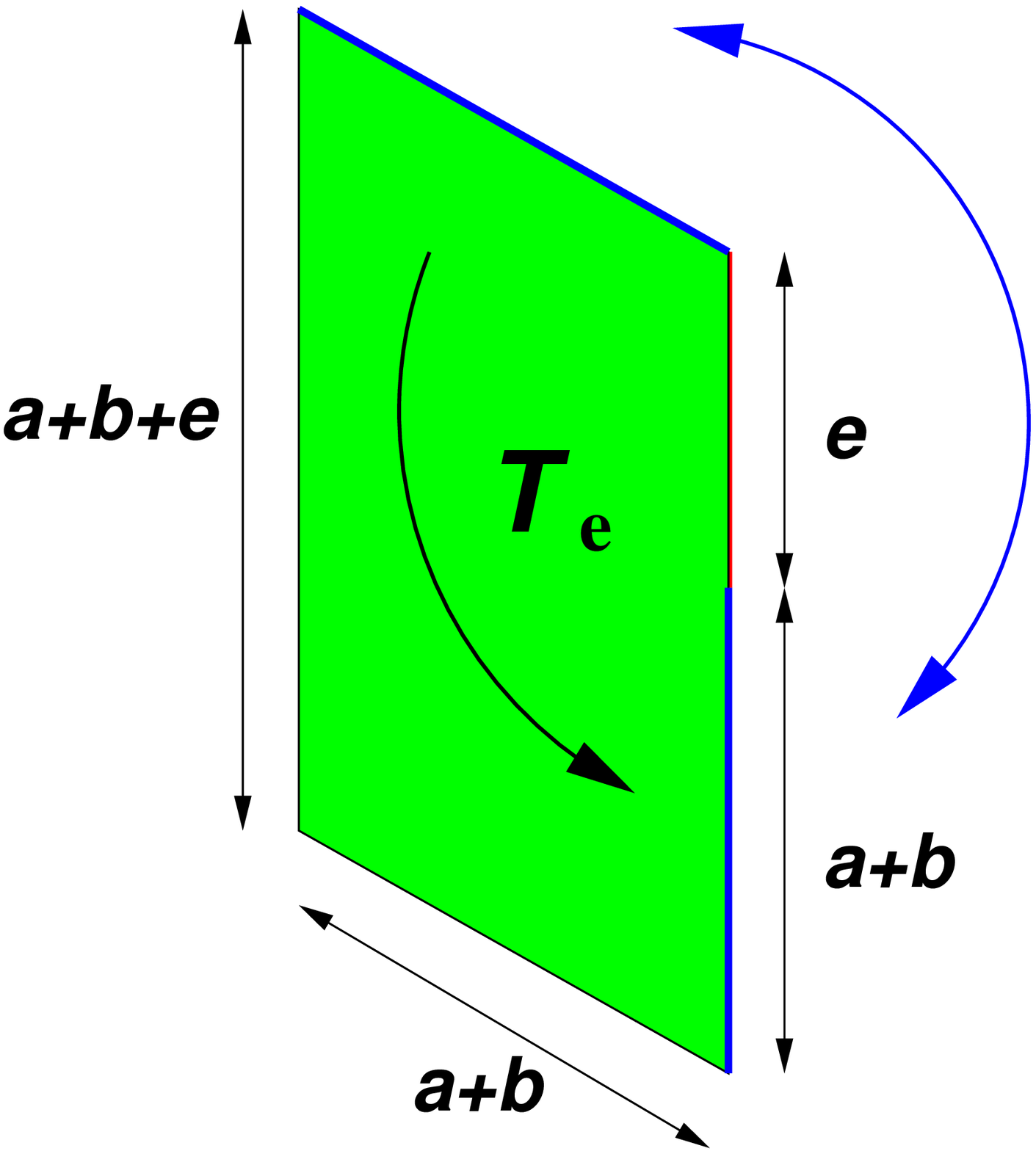}{4.5cm}
\figlabel\fivelinesht

The first class is the so-called half-turn symmetric FPL (HTSFPL), 
i.e.\ configurations
which are invariant by rotation of $\pi$. Naturally,
the connectivity pattern itself must be half-turn symmetric, so that we must choose
it to be of the form $(a,b|e|b,a)$. The domain is of the form of Fig.~\fivelinesht,
and we find that

\eqn\NHT{
N^{HT}(a,b|e|a,b)= \det\left(I+\mu\, T_e(a+b)\right)|_{\mu^b}
}
where $T_e(a+b)=T(a+b)X_e(a+b)$ has its entries given by Eq.~\newmat\ with $1\le i,j\le a+b$.

Note in particular that for $e=0$ we recover the
sequence A045912 of \Sloane\ that was already mentioned in section 3.1 (Eq.~\simply); and
by summing at fixed $n=a+b$, the total number of dimers
$\det(I+T(n))$ (Eq.~\exone; sequence A006366 of \Sloane) which is 
the number of cyclically symmetric plane partitions in the $n$-cube.
This number also coincides with
the ratio of numbers of $2n\times 2n$ half-turn symmetric ASM (HTSASM) and of $n\times n$ ASM. 
One now finds an indirect proof of the latter fact, via the
Razumov--Stroganov conjecture for HTSFPL and FPL. Indeed, there is a projection
that sends the ground state eigenvector counting FPL onto the one for HTSFPL.
All connectivity patterns that contribute to 
$N^{HT}(a,b| 0| a,b)$ are projected\foot{Recall that
a HT-symmetric pattern of size $2n$ is projected onto a pattern
of size $n$ by cutting the disk into 2 equal sized pieces in any way such
that it does not cut an arch, taking one half-disk and gluing it back
together into a disk.} onto the trivial pattern of $n$ arches
for FPLs of size $n$; there is only one corresponding FPL, and
therefore the sum of such contributions must be related to the relative
normalization of the ground state eigenvectors, which can be obtained
by taking the sum of all components, i.e.\ which is equal
to the ratio of numbers of $2n\times 2n$ HTSASM 
and of $n\times n$ ASM.

\fig{Domains of rhombus tilings for the (i) $(a,b|e|b,a)$ VSFPL and 
(ii) $(a,1,a|e|a,1,a)$ HVSFPL configurations.
In each case there are exactly $a$ non-intersecting paths (only one is depicted).}{%
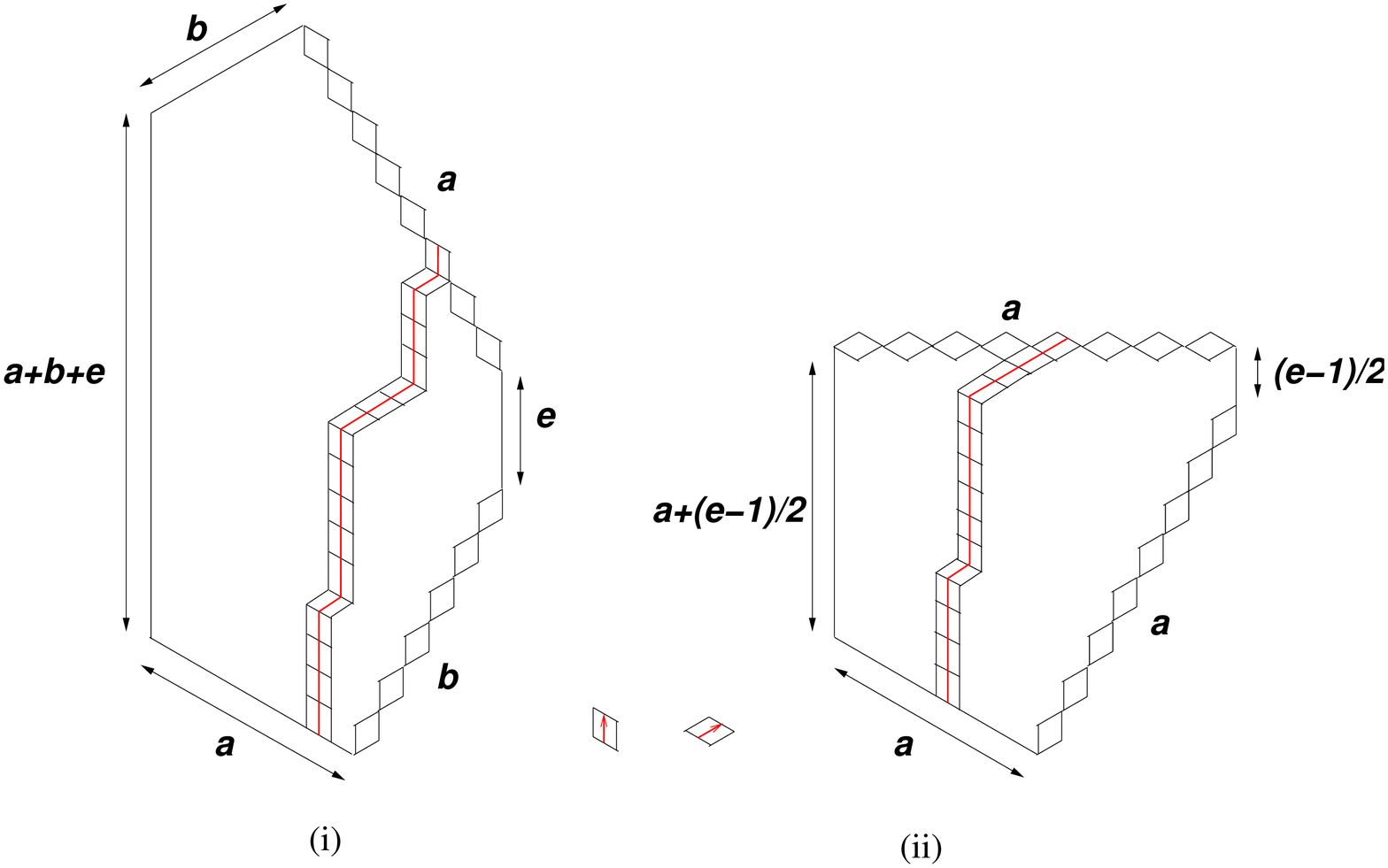}{12cm}
\figlabel\fivelinesvs

Next we consider the vertically symmetric FPL (VSFPL), which
are invariant by reflection with respect to the vertical axis, and
only exist for odd sizes. The connectivity is now $(a,b|e|b,a)$, $e$ odd, and
the domain is of the form of Fig.~\fivelinesvs~(i). There are exactly $a$ non-intersecting
paths, and it is convenient to redefine them as coming out of the side of length $a$ and
propagating to the opposite side as depicted. As in 
section (3.4), we apply the reflection principle to compute the number of paths 
in the presence of a ``wall'', 
and then apply the LGV formula:
\eqn\NV{
N^{V}(a,b| e| b,a)=\det_{i,j=1\ldots a}\left(
{2b+e+j-1\choose b-j+i}-{2b+e+j-1\choose b-j-i+1}\right)\ .
}

Finally, the vertically and horizontally symmetric FPL (HVSFPL) are invariant
under reflections with respect to both axes. Here the
sequence of arches must be slightly modified in order to accomodate the additional
symmetry; with the insertion of two single arches, the connectivity pattern becomes
$(a,1,a|e|a,1,a)$ and
the domain is of the form of Fig.~\fivelinesvs~(ii). We finally find
\eqn\NHV{
N^{HV}(a,1,a| e| a,1,a)=\det_{i,j=1\ldots a}\left(
{2a+(e-1)/2-j+1\choose a-2j+i+1}-{2a+(e-1)/2-j+1\choose
a-2j-i+2}\right)\ .
}

\subsec{Plane partitions: $q$-decoration}
When we view the rhombus tilings of a hexagon as plane partitions, we may introduce yet another
catalytic variable, namely a weight $q$ per unit cube in the PP. Many of the above results may thus be 
``$q$-decorated"  whenever there exists a PP interpretation.
This is for instance the case for the total number of PP in a box of size
$a\times b\times c$, which gives a $q$-MacMahon formula. 
To express it, we introduce $q$-deformed corner transfer matrices $w,t$
that keep track of the numbers of boxes in the PP picture.
This leads to
\eqn\qW{w_{ij}=q^{{1\over 6}+{i-1\over 2}+{(j-1)^2\over 2}} \left[\matrix{ i-1\cr
j-1}\right]_q}
where we have used $q$-binomials ($\left[\matrix{ a\cr b}\right]_q=(1-q^{b+1})(1-q^{b+2})\ldots
(1-q^a)/((1-q)(1-q^2)\ldots(1-q^{a-b}))$), and where the prefactors
are ad-hoc to take care of boundaries.
Similarly, we have $t=ww^t$ reading
\eqn\qT{t_{ij}= q^{{1\over 3}+{i+j-2\over 2}} \left[\matrix{ i+j-2 \cr
i-1}\right]_q \ .}
With this definition, we get the $q$-deformation of the MacMahon formula \mac\
\eqn\qdefmac{ N(a,b,c; q)=\det(I+t(a,b)t(b,c)t(c,a))=\prod_{i=1}^a\prod_{j=1}^b\prod_{k=1}^c
{1-q^{i+j+k-1}\over 1-q^{i+j+k-2}} }
for the generating function $N(a,b,c; q)$ for PP in a box $a\times b\times c$ with a weight $q$
per unit cube.
Combining this with eq.\dnumber, we also get the generating function $N(a,b,c; q,\mu)$
for PP in a box $a\times b\times c$ with a weight $q$
per unit cube, and with a weight $\mu$ per De Bruijn loop in the tiling picture
\eqn\maqdef{ N(a,b,c; q,\mu)=\det(I+\mu \,t(a,b)t(b,c)t(c,a))\ .}

With the matrix $t$, we have also access to the generating function $N^{CS}(a; q,\mu)$
for the numbers of cyclically symmetric plane partitions
in a box $a\times a\times a$, weighted by $q$ per unit cube, and $\mu$ per winding De Bruijn loop, namely 
\eqn\matTCS{N^{CS}(a; q,\mu)=\det(I+\mu\, t(a))}
provided $q$ is replaced by $q^3$ in the definition of $t$ (as the total number of 
boxes is three times that in any of the three identical copies of the parallelogram that form 
the total hexagon).

\fig{Vertically symmetric plane partitions in a box $a\times a\times a$
corresponding to the tiling of a glued half-hexagon, for $a=2$.
For each partition, we have indicated the contribution to the generating function
$N^{HH}(2; q,\mu)=1+q \mu+q^3\mu+q^5\mu+q^7\mu+q^8\mu^2$.}{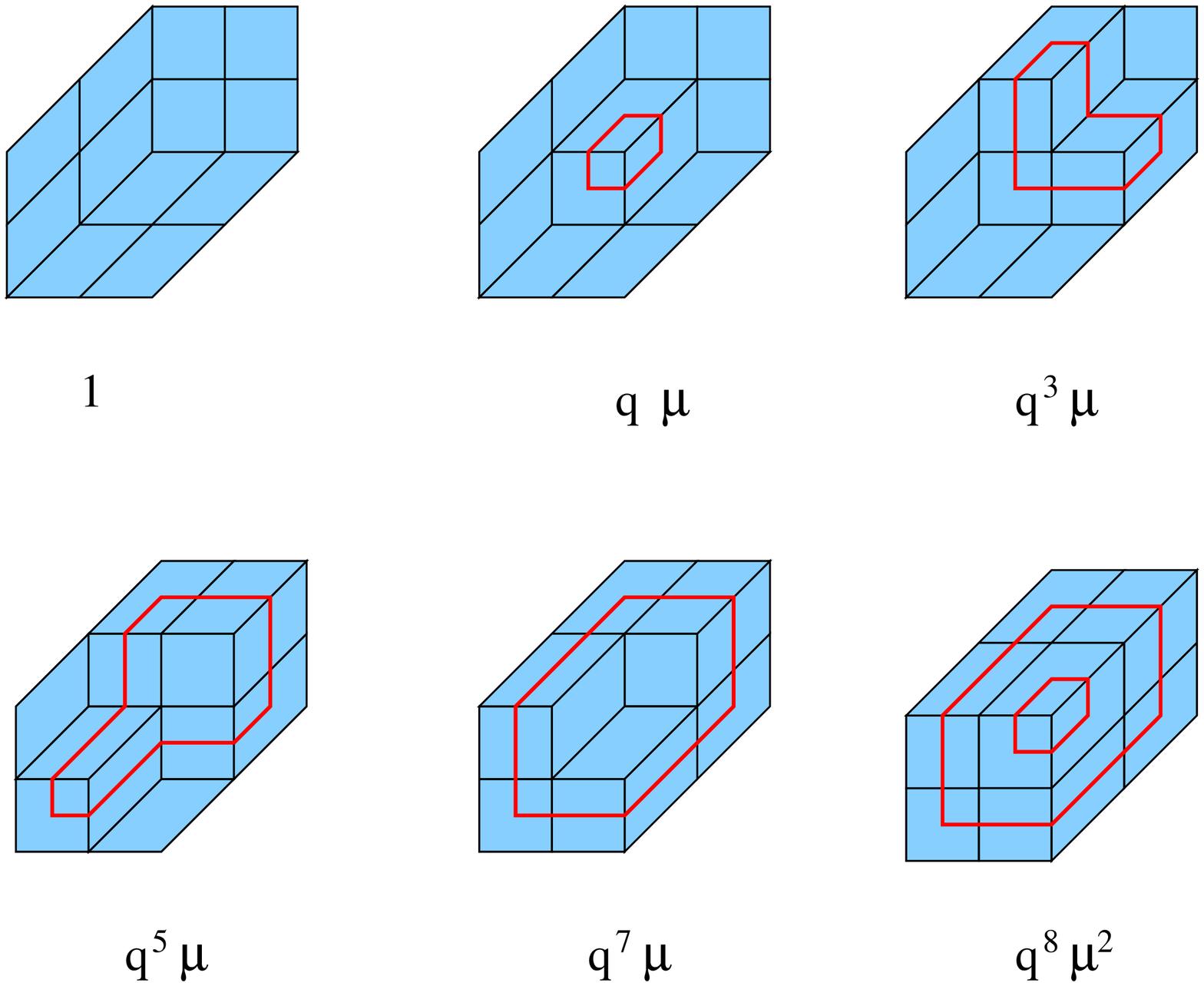}{8.cm}
\figlabel\halfturn

Finally, we get the generating function $N^{HH}(a; q,\mu)$
for the numbers of vertically symmetric partitions corresponding to the tiling of
the glued half-hexagons of section 3.2, with an additional weight $q$ per unit cube 
and $\mu$ per winding De Bruijn loop
\eqn\matTHT{N^{HH}(a; q,\mu)=\det(I+\mu \,w(a)w^t(a)w(a))}
provided $q$ is replaced by $q^2$ in the definition of $w$ (as the plane partition is
obtained by completing the (opened) half-hexagon with its reflection.
These partitions are represented in Fig.~\halfturn\ for $a=2$. 
We may also write a $q$-deformed
version of the Poincar\'e polynomial \poinca\ for 2-step nilpotent Lie algebras of rank $a$:
\eqn\qpoinca{ P_a(u;q)=\det\big( I+{1\over \sqrt{q}}\theta w(a)w^t(a)w(a)\big) }
with the matrix $\theta$ as in \thetentries. The latter corresponds presumably to
a sum over Schur functions of the form $\sum_{Y=Y^T} s_Y(x_q)u^{|Y|}$,
with the specialization $x=x_q\equiv \{1,q,q^2,\ldots,q^{a-1}\}$. The coefficients
of $P_a(u;q)$ as a function of $u$ are $q$-deformed dimensions
of homology spaces, and await a good interpretation in the Lie algebraic context.
For illustration, we have
\eqn\pnex{\eqalign{ 
P_3(u;q)&=1+u(1+q+q^2)+u^3q^2(1+q)^2(1+q^2)+u^4q^3(1+q^2)(1+q+q^2)\cr
&+u^5 q^5(1+q^2)(1+q+q^2)+u^6 q^6(1+q)^2(1+q^2)+u^8q^{10}(1+q+q^2)+u^9
q^{12}\ . \cr}}


\newsec{Conclusion}
In this paper, we have provided expressions for the numbers of rhombus tilings of certains
regions of the plane with some gluing conditions. Sometimes these expressions were directly
determinants, sometimes they were coefficients in the expansion of a determinants in powers of a variable,
in a way similar to grand canonical vs canonical partition functions. We used transfer matrix techniques,
in analogy with Baxter's corner transfer matrix.
We have applied these ideas to the
computation of the numbers of FPL for a fairly general class of connectivities, with and without
additional symmetries. Various issues remain open.

\subsec{FPL and rhombus tilings}

In view of sect. 4.2 and 4.3, one may wonder whether there exists a
deeper  connection between FPL \conf s with prescribed connectivities
and rhombus tilings of possibly cut and/or glued domains of the plane.

The answer to this question must be subtle as was already observed 
in the case of four sets of nested lines \DFZ: Indeed in that case,
only a few  rotated versions (in the sense of Wieland \Wie) of
the FPL counting problem allowed for a straightforward bijection with 
rhombus tilings of the domains of figure 11. In the general case, 
we must first optimize the Wieland rotation in order to attack the 
problem.

Another relation to rhombus tiling, though very hypothetical, 
would first involve finding a bijection between FPL/ASM 
and totally symmetric self-complementary plane partitions (TSSCPP), 
themselves reducible to the rhombus tilings of triangles 
with free boundary conditions along one side.
The number of TSSCPP in a box of size $2a\times 2a\times 2a$
was shown to match that of $a\times a$ FPL/ASM, but no natural
bijection is available yet between the two sets \BIBLE.
If we knew such a bijection, we could characterize among TSSCPP
the rhombus tilings corresponding to FPL  \conf s with fixed
connectivities, thus provide an answer to the above question.
This answer, however, would not coincide with that of sect. 4.2
in the case of four sets of lines separated by a fifth one, as 
the domain considered do not match in any simple way. 

Answering this question remains an interesting challenge.

\subsec{Asymptotic enumeration}
A by-product of our work has been to provide  us with new data for FPL \conf\ numbers, 
either by explicit  determinant formulae, or just numerical. It is tempting to 
try to extract the asymptotic behavior for large \conf s of a given type. 
This is in particular the case for the \conf s denoted $(\ )_r^p$ in \MNdGB,
i.e. made of $p$ sets of $r$ arches, for which we conjecture that
$p$ large, $r$ fixed, or vice versa (and $n=p\, r$)
\eqn\conjasymp{\log\#\left((\ )_r^p\right)
\approx \kappa r^2 ((p-1)^2-1)  + {\cal O}(\log n)}
with the same $\kappa={1\over 2}\log{ 27\over 16}$ that appears in the 
asymptotics of $A_n$, 
\eqn\asymptAn{ A_n\approx n^{-23/36} e^{\kappa n^2}.}

$$\hskip-4mm
\matrix{p\backslash r & 1 & 2&3&4 & \  &   \alpha_p \cr
2 : &  1 & 1 & 1& 1 & & 0\cr
3 : &2 & 20 & 980 & 232848& & 3/2 \cr
4 : &7 &  3504 &  118565449 & 266866085641550  & & \sim 4 \ (*)\cr
5 : &42 & 5100260 & 1637273349805800 & \cdots & & \sim 15/2 ? \cr
6 : &429 &60908609580 &  \cdots \cr
7 : &7436&5939300380261111&\cdots\cr
8:  &218348 &4717858636573174999768&\cdots\cr
& \cr
\beta_r & {1\over 2} & \sim 2 ? & \sim {9\over 2}? \cr
} $$

\centerline{\bf Table 1. Numbers of FPL configurations 
$()_r^p$ with $p$ sets of $r$ arches}
\par\begingroup\parindent=0pt\leftskip=1cm\rightskip=1cm\parindent=0pt
\baselineskip=11pt
\midinsert
{ ($*$) see the text for justification. In each column, as $p\to \infty$, 
there is good evidence that the numbers behave as $\left({27\over 16}\right)^{\beta_r p^2}$, 
with numbers $\beta_r$ as shown; likewise, along each row, 
$r\to \infty$, behavior as $\left({27\over 16}\right)^{\alpha_p  r^2}$, with $\alpha_p$ as
shown. The expression in \conjasymp\ is consistent with these data. }
\vskip 12pt\par
\endinsert\endgroup\par
\goodbreak
\global\advance\figno by1

The conjecture \conjasymp\ agrees with some  particular
cases. For $r=1$,  all $p=n$,  we have  \conf s
with $n$ simple arches, whose number is
$A_{n-1}$, as conjectured in \BdGN, and whose asymptotics is thus
given by \asymptAn\ (with $n\mapsto n-1$).  For 
$p=3$, $r=n/3$, we have the MacMahon formula, the asymptotics of which 
is easy to calculate. Our conjecture is also supported by
the observation that the number of 4-arch \conf s $(\ )_r^4$, $r=n/4$,
is given by the middle term in the expansion of $\det (x+T^2)$,
with $T$ the Pascal matrix \Telem,  and
is in fact the dominant coefficient in that polynomial in $x$.
Then Mitra-Nienhuis \MNosc\ 
conjecture on $\det(i 1+T)\sim (A^{\rm HT}(L=2r))^2$ together with
the known asymptotic
behavior of half-turn symmetric ASM,  
$A^{\rm HT}(L) \sim \left({27\over 16}\right)^{L^2/4}$,
gives $\log \#(\ )_r^4\sim {1\over 2} {8 r^2 } \log{27\over 16}$.
Last but not least, our conjecture \conjasymp\ is well supported by
numerical data, as shown in Table 1.


\bigskip\bigskip\bigskip
\nind{\bf Acknowledgments} 

Many thanks to Gleb Koshevoy, Christian \Kr\ and Nikolai Reshetikhin
for discussions and suggestions. 
This work was partly supported by the  TMR network EUCLID
 HPRN-CT-2002-00325.

This work is dedicated to our friend Pierre van Moerbeke
on the occasion of his sixtieth birthday, and we are
happy to wish him many more years of exciting research.


\listrefs

\end